%% file: main.tex
\documentclass[conference, letterpaper]{IEEEtran}

\usepackage{multirow}    
\usepackage{pifont}      
\usepackage{color}       
\usepackage{xspace}      
\usepackage{flushend}     
\usepackage{graphicx}
\usepackage{subcaption}
\usepackage{url}
\usepackage{rotating}
\usepackage{paralist}

\usepackage{enumitem}
\usepackage{soul}
\usepackage[linesnumbered,ruled]{algorithm2e}
\makeatletter
\renewcommand{\@algocf@capt@plain}{above}
\makeatother 
\usepackage{algorithmicx}  
\usepackage{algpseudocode}  
\usepackage{amsmath,amsthm} 
\algnewcommand{\LineComment}[1]{\State \(\triangleright\) #1}
\usepackage{tikz}
\usepackage{xcolor}
\usepackage[normalem]{ulem}

\begin{document}

\newcommand{\mypara}[1]{\vspace{2pt}\noindent\textbf{{#1: }}}
\newcommand{\eat}[1]{}  
\newcommand{\upperperi}{PeriHammer\xspace}
\newcommand{\uppertele}{TeleHammer\xspace}
\newcommand{\upperpt}{PThammer\xspace}
\newcommand{\authcomment}[3]{\textcolor{#3}{#1 says: #2}}
\newcommand{\zhi}[1]{\authcomment{zhi}{#1}{blue}}
\newcommand{\yueqiang}[1]{\authcomment{yueqiang}{#1}{red}}
\newtheorem{definition}{Definition}

\title{\emph{\uppertele}: A Formal Model of Implicit Rowhammer}

\author{
\IEEEauthorblockN{Zhi Zhang$^{*,1,2}$, Yueqiang Cheng$^{*,3}$, Dongxi Liu$^2$, Surya Nepal$^2$, and Zhi Wang$^4$}
\IEEEauthorblockA{$^*$ Both authors contributed equally to this work \\ $^1$ University of New South Wales \\ $^2$ Data61, CSIRO, Australia \\ $^3$ Baidu Security \\ $^4$ Florida State University, America \vspace{-0.2cm}}
}
\maketitle

\begin{abstract}
The rowhammer bug is to frequently access hammer rows to induce bit flips in their adjacent victim rows, allowing an attacker to gain privilege escalation or steal private data. A key requirement of \emph{all} existing rowhammer attacks is that an attacker must have access to at least part of an exploitable hammer row. We refer to such rowhammer attacks as \upperperi. The state-of-the-art software-only defenses against \upperperi attacks is to make the exploitable hammer rows beyond the attacker's access permission.

In this paper, we question the necessity of the above requirement and propose a new class of rowhammer attacks, termed as \uppertele. It is a paradigm shift in rowhammer attacks since it crosses privilege boundary to stealthily rowhammer an inaccessible row by implicit DRAM accesses. Such accesses are achieved by abusing inherent features of modern hardware and/or software.
We propose a generic model to rigorously formalize the necessary conditions to initiate \uppertele and \upperperi, respectively. Compared to \upperperi, \uppertele can defeat the advanced software-only defenses, stealthy in hiding itself and hard to be mitigated.

To demonstrate the practicality of \uppertele and its advantages, we have created a \uppertele's instance, called \upperpt, which leverages the address-translation feature of modern processors. We observe that a memory access from user space can induce a load of a Level-1 page-table entry (L1PTE) from memory and thus hammer the L1PTE once, although L1PTE is not accessible to us. To achieve a high enough hammering frequency, we flush relevant TLB and cache \emph{effectively} and \emph{efficiently}. To this end, we demonstrate \upperpt on three different test machines and show that it can cross user-kernel boundary and induce the first bit flips in L1PTEs within 15 minutes of double-sided hammering. 
\upperpt does not require the \emph{superpage} system setting, and works on Ubuntu Linux. We have exploited \upperpt to defeat advanced software-only rowhammer defenses in default system setting.
\end{abstract}

\input{intro.tex}
\input{bkgd.tex}
\input{overview.tex}
\input{eva.tex}
\input{discuss.tex}

\input{conclusion.tex}
\bibliographystyle{plain}
\bibliography{main}
\appendix
\input{appendix.tex}
\end{document}

%% file: intro.tex
\section{Introduction}\label{sec:intro}
\begin{figure}
\centering
\includegraphics[width=\columnwidth]{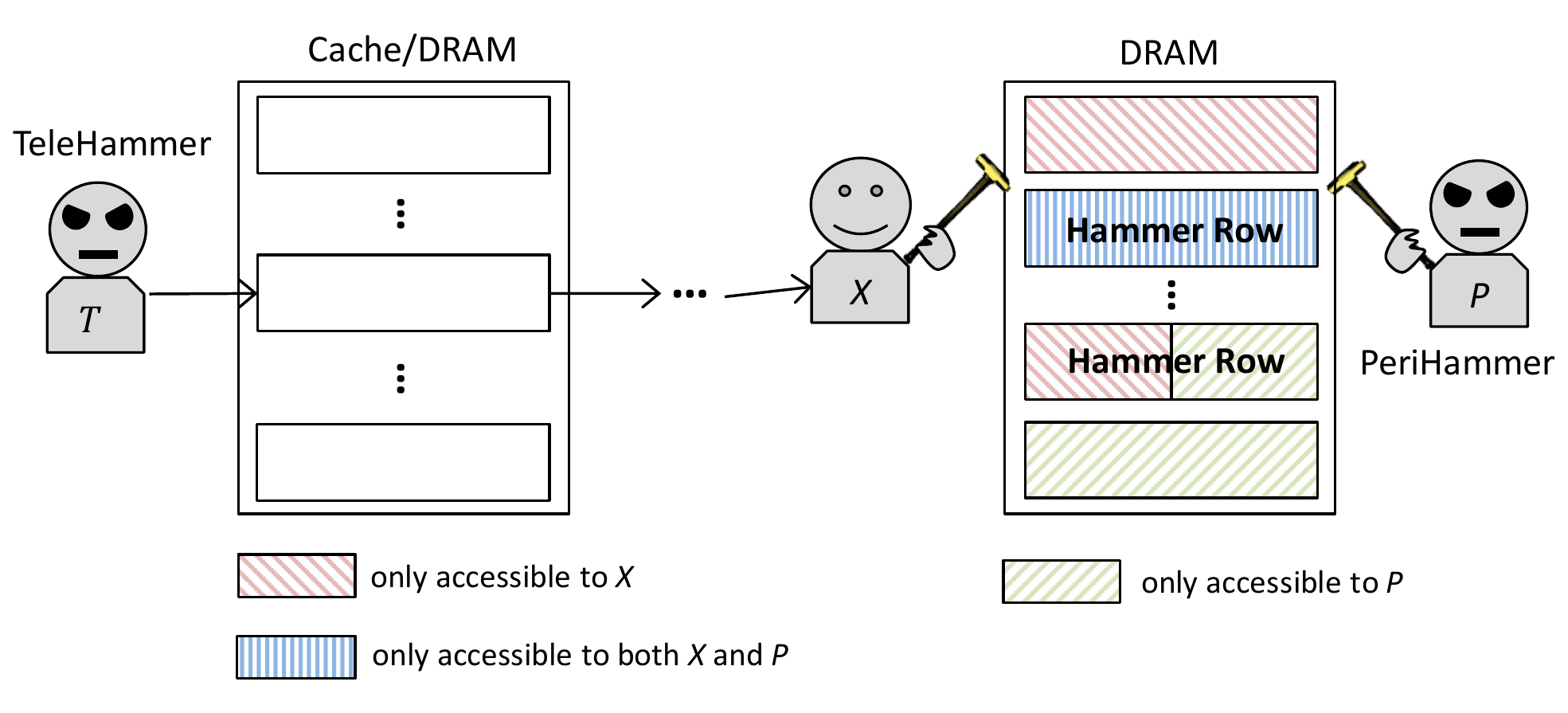}
\caption{
\emph{\uppertele} and \emph{\upperperi}. 
In \emph{all} existing rowhammer attacks (i.e., \upperperi), an attacker \emph{P} requires exploitable hammer rows to be at least partially accessible. An attacker \emph{T} in \emph{\uppertele} removes this requirement by tricking a benign entity \emph{X} (e.g., system call handler) to implicitly hammer the exploitable hammer rows. Compared to \upperperi, the across-privilege-boundary property makes \uppertele stealthy and hard to be mitigated, which is the main difference between \uppertele and \upperperi.}
\label{fig:indirect_hammer}
\end{figure}

In 2014, Kim et al.~\cite{kim2014flipping} discovered an infamous software-induced hardware fault, the so-called ``rowhammer'' bug.  
Specifically, frequent accessing the same addresses in two DRAM (Dynamic Random Access Memory) rows (i.e., hammer rows) can cause bit flips in an adjacent row (i.e., a victim row). If sensitive structures, such as page tables, are placed onto the victim row, an adversary can corrupt the structures by exploiting adjacent hammer rows although she has no access to the structures. As such, the bug can be exploited to break MMU-based domain isolation between different security domains (e.g., user and kernel) without software vulnerabilities, enabling a powerful class of attacks targeting DRAM-based systems. 
The attacks are so hazardous that they can either gain the privilege escalation~\cite{seaborn2015exploiting, gruss2016rowhammer, bosman2016dedup, frigo2018grand, tatar2018throwhammer,gruss2017another,van2016drammer, xiao2016one} or steal the private data~\cite{razavi2016flip,bhattacharya2016curious, kwong2020rambleed}.
To exploit the bug, all existing rowhammer attacks require access to at least part of an \emph{exploitable} hammer row (a hammer row is exploitable when part of it is sensitive~\cite{kwong2020rambleed} or its adjacent victim row is sensitive~\cite{seaborn2015exploiting} ) as shown in Figure~\ref{fig:indirect_hammer}.  As their access to the hammer row is legitimate and conforms to the privilege boundary enforced by MMU, we term such attacks as \upperperi.

To defeat \upperperi-type attacks, numerous hardware and software based mitigation techniques have been proposed. 
As hardware based mitigation require DRAM updates or upgrade and cannot be backported, recent software only defenses including CATT~\cite{brasser17can}, RIP-RH~\cite{bock2019rip} and CTA~\cite{wu2018CAT} are practical for bare-metal systems. These defenses in common enforce DRAM-based memory isolation at different granularity to deprive the attackers of access to the exploitable hammer rows. Take CTA~\cite{wu2018CAT} as an example, it primarily isolates DRAM memory of page tables such that all exploitable hammer rows that can flip bits in page tables are beyond the privilege boundary of unprivileged attackers.


\eat{
\begin{figure}
\centering
\includegraphics[scale=0.45]{image/PeriHammer.pdf}
\caption{
\upperperi. \emph{All} existing rowhammer attacks require at least part of a hammer row accessible to an attacker with the purpose of either gaining privilege escalation or stealing private data.} 
\label{fig:normal_hammer}
\end{figure}
}

\mypara{\uppertele}
in this paper, we introduce a paradigm shift in rowhammer attacks through a new class of rowhammer attacks, called \uppertele. As shown in Figure~\ref{fig:indirect_hammer}, an attacker \emph{T} in \uppertele tricks a benign entity to implicitly hammer the exploitable rows that are inaccessible. This cross-privilege-boundary property thus eliminates the above key requirement of \upperperi.
Essentially, \uppertele abuses built-in features of modern hardware and/or software (such as speculative execution, system call handler, etc.) and the entity is either hardware such as processor or software such as system call handler. For instance, an unprivileged attacker can exploit speculative execution to cross user-kernel privilege boundary and access kernel memory~\cite{kocher2018spectre}. If each access occurs frequently within DRAM, then the accessed kernel memory might be vulnerable to the rowhammer bug.

To exploit the rowhammer bug, \uppertele has the following challenges to overcome. First, 
an unprivileged attacker such as \emph{T} in Figure~\ref{fig:indirect_hammer} should find one feature of either hardware or software to identify a path (e.g., solid line with an arrow in Figure~\ref{fig:indirect_hammer}) to the inaccessible and exploitable hammer rows. Second, \emph{T} should have an approach to effectively specify the path each time when tricking \emph{X} into hammering the hammer rows. Third, both specifying and hammering operations should be efficient enough so as to trigger the rowhammer bug (see details in Section~\ref{sec:telehammer}). 

We propose a generic formal model to rigorously formalize the above challenges. Also, the model can be used to formalize \upperperi and show that \upperperi is a special case of \uppertele. \uppertele exhibits the following advantages over \upperperi. 
\begin{itemize}[noitemsep, topsep=2pt, partopsep=0pt,leftmargin=0.4cm]
    \item \uppertele can defeat the aforementioned advanced software-only defenses~\cite{brasser17can,bock2019rip,wu2018CAT}, since it uses cross-privilege-boundary rowhammer and eschews the aforementioned critical requirement of \upperperi. 
    \item \uppertele is stealthy, since it hammers a hammer row not by itself but using a target domain (e.g., kernel), making it hard to trace the real culprit.
    \item \uppertele is difficult to defend against, since abundant instances can be derived by its design. A countermeasure against specific instances cannot defeat \uppertele. 
\end{itemize}
  
\mypara{\upperpt}
to demonstrate the practicality of \uppertele, we create a working instance, called \upperpt, that satisfies the formal model of \uppertele. Specifically, 
we observe that a memory access triggers the address translation in modern OSes on x86-64 microarchitecture. In response to the memory access, the processor first searches Translation-lookaside Buffer (TLB) to check if a corresponding physical address exists. If the search fails (i.e., a TLB miss), then the processor searches paging structure that hosts a partial address mapping of different page-table levels~\cite{barr2010translation}. If another miss occurs, it fetches four-level page-table entries (PTEs) from CPU cache, otherwise DRAM memory. Fetching PTEs from memory causes hammering the PTEs once. Intuitively, the four-level PTEs will be hammered if TLB, paging structure and cache are flushed out effectively and efficiently. 

\upperpt abuses the above address-translation feature to target Level-1 PTEs (L1PTEs). It requires \emph{effectively} and \emph{efficiently} flushing corresponding address mappings from TLB and relevant L1PTEs from cache.
However, it is not authorized to explicitly perform such flushes by the means of instructions.
Instead, \upperpt first prepares a complete pool of eviction sets for TLB and cache, respectively, which can be used to implicitly flush a target entry from TLB and a target memory line from cache.
From both TLB and cache eviction pools, \upperpt respectively selects two eviction sets for subsequent double-sided hammering. 
Note that the pool preparation for either TLB or cache is a one-off cost and can be done at the beginning of \upperpt. 
We launch \upperpt as an unprivileged process in two system settings (i.e., \emph{superpage} system setting is disabled by default or enabled) on three different machines. The experimental results indicate that \upperpt  is able to cross the user-kernel boundary and induce the first bit flip in L1PTEs within 15 minutes of double-sided hammering in either setting. 
Furthermore, we develop \upperpt-based exploit to defeat all the aforementioned practical defenses (see details in Section~\ref{sec:eva}).



Note that \upperpt is not the only working instance of \uppertele, and there should exist other instances that
also leverage built-in system features to achieve the frequent implicit DRAM accesses. 
We discuss potential instances in Section~\ref{sec:dis}.
 

\eat{

\emph{First}, it removes the necessity of implicitly flushing paging structure (see Section~\ref{sec:pthammer}), which requires the publicly-unknown mapping between a virtual address and its paging structure indexing. 
\emph{Second}, it only requires flushing TLB and cache of L1PTEs, thus greatly improving the hammer efficiency.
\emph{Third}, L1PTE based \upperpt consumes much less memory, making itself stealthy. We apply previous works~\cite{seaborn2015exploiting,cheng2018still} to create a certain number of only Level-1 page-table pages for hammering.
\emph{Last}, we can gain access to the whole system memory if we can exploit bit flips in one L1PTE.
}


\mypara{Summary of Contributions}
the main contributions of this paper are as follows:
\begin{itemize}[noitemsep, topsep=2pt, partopsep=0pt,leftmargin=0.4cm]
    \item All previous rowhammer exploits (i.e., \upperperi) require access permission to exploitable hammer rows. In contrast, we propose a new class of rowhammer attacks, called \uppertele, that proposes cross-privilege-boundary rowhammer and voids the critical requirement. 
    \item We present a generic model to formally define necessary conditions to launch \uppertele and \upperperi, respectively. Based on the model, we summarize three advantages of \uppertele over \upperperi.
    \item We propose an instance of \uppertele, called \upperpt, that leverages the address-translation feature to cross the user-kernel boundary and induce rowhammer bit flips in Level-1 page tables.
    \item We evaluate \upperpt in two system settings (i.e., \emph{superpage} is either inactive by default or active) on three different machines, which indicate that \upperpt can cross the user-kernel boundary and induce the first bit flip in Level-1 page table entries within 15 minutes of double-sided hammering, and defeat the aforementioned advanced software-only defenses in default system setting.
\end{itemize}

\eat{
The rest of the paper is structured as follows.
In Section~\ref{sec:bkgd}, we briefly introduce the background information and summarize related works. 
In Section~\ref{sec:overview}, we present a formal model of \uppertele and talk about how to instantiate \uppertele in detail.
Section~\ref{sec:eva} evaluates \upperpt thoroughly. In Section~\ref{sec:dis}, we discuss how to compromise ZebRAM~\cite{konoth2018zebram}, a defense for virtualization systems, shed light on other possible instances of \uppertele and discuss possible mitigation against \uppertele and \upperpt. We conclude this paper in Section~\ref{sec:conclusion}.
}


%% file: bkgd.tex
\section{Background and Related Work}\label{sec:bkgd}
In this section, we first introduce CPU cache, Translation-lookaside Buffer (TLB), Dynamic Random-access Memory (DRAM) and then the rowhammer bug as well as its attacks.

\subsection{CPU Cache}\label{sec:cache}
In commodity Intel x86 micro-architecture platforms, there are three levels of CPU caches. Among all levels of caches, the first level of cache (i.e., L1 cache) is closest to CPU.
L1 cache has two types of caches, i.e., L1D caching data and L1I caching instructions. The second level of cache, L2, is unified caching both data and instructions. Similar to L2, the last-level cache (LLC) or L3, is also unified.  
Generally speaking, cache of a specific level is set-associative and it consists of \emph{S} sets. Each set contains \emph{L} lines and data or code can be cached in any line of the set; this is referred as an \emph{L}-way set-associative cache set. For each cache line, it stores \emph{B} bytes. Thus, the overall cache size of that level will be $\emph{S} \times \emph{L} \times \emph{B}$.  

When an accessed variable is stored in a cache set, Intel micro-architectures use its virtual or physical address to decide its corresponding cache set of a specific cache level. For instance, L1 cache set is indexed using bits 6 to 11 of a virtual address. For L3, its indexing scheme is more complicated. In contrast to L1 and L2 that are private to a physical core, L3 is shared among all cores. So L3 cache is firstly partitioned into slices, and one slice serves one core with a higher priority. For each slice, it is further divided into cache sets as mentioned above. As such, some physical-address bits are XORed to decide a slice, and some bits (bits 6 to 16) are XORed to index a cache set~\cite{liu2015last}.

\subsection{Translation-lookaside Buffer}
Translation Lookaside Buffer (TLB) has two levels. 
The first-level (i.e., L1), consists
of two parts: one that caches translations for code pages, called L1 instruction TLB (L1 iTLB), and the other that caches translations for data pages, called L1 data TLB (L1 dTLB). The second level TLB (L2 sTLB) is larger and shared for translations of both code and data.
Similar to the CPU cache above, the TLB at each level is also partitioned into sets of ways. One way is a TLB entry that stores one address mapping between a virtual address and a physical address. 

Note that a virtual address (VA) determines a TLB set of each level. Although there is no public information about the mapping between the VA and the TLB set, it has been reverse-engineered on quite a few Intel commodity platforms~\cite{gras2018translation}.

\subsection{Dynamic Random-access Memory}
Main memory of most modern computers uses Dynamic Random-access Memory (DRAM). Memory modules are usually produced in the form of dual inline memory module, or DIMM, where both sides of the memory module have separate electrical contacts for memory chips. Each memory module is directly connected to the CPU's memory controller through one of the two channels. Logically, each memory module consists of two ranks, corresponding to its two sides, and each rank consists of multiple banks. A bank is structured as arrays of memory cells with rows and columns. 

Every cell of a bank stores one bit of data whose value depends on whether the cell is electrically charged or not. A row is a basic unit for memory access. Each access to a bank ``opens'' a row by transferring the data in all the cells of the row to the bank's row buffer. This operation discharges all the cells of the row. To prevent data loss, the row buffer is then copied back into the cells, thus recharging the cells. Consecutive access to the same row will be fulfilled by the row buffer,  while accessing another row will flush the row buffer.

\subsection{Rowhammer Overview}\label{sec:hammeroverview}

\mypara{Rowhammer bugs}
Kim et al.~\cite{kim2014flipping} discovered that current DRAMs are vulnerable to disturbance errors induced by charge leakage. In particular,  their experiments have shown that frequently opening the same row  (i.e., hammering the row) can cause sufficient disturbance to a neighboring row and flip its bits without even accessing the neighboring row. Because the row buffer acts as a cache, another row in the same bank is accessed to flush the row buffer after each hammering so that the next hammering will re-open the hammered row, leading to bit flips of its neighboring row. 

\mypara{Hammering techniques} generally speaking, there are three techniques regarding hammering a vulnerable DRAM.

\emph{Double-sided hammering:} it is the most efficient technique to induce bit flips. Two adjacent rows of a victim row are hammered simultaneously and the adjacent rows are called hammer rows~\cite{kim2014flipping}.

\emph{Single-sided hammering:}
Seaborn et al.~\cite{seaborn2015exploiting} proposed a single-sided hammering by randomly picking multiple addresses and just hammering them with the hope that such addresses are in different rows within the same bank. 

\emph{One-location hammering:}
one-location hammering ~\cite{gruss2017another} randomly selects a single address for hammering. It exploits the fact that advanced DRAM controllers employ a more sophisticated policy to optimize performance, preemptively closing accessed rows earlier than necessary. 

\mypara{Key requirements} the following requirements are needed by \upperperi-based attacks to gain either privilege escalation or private information. 

\emph{First}, 
CPU cache must be either flushed or bypassed. 
It can be invalidated by instructions such as \texttt{clflush} on x86. In addition, conflicts in the cache can evict data from the cache since cache is much smaller than the main memory. Therefore, to evict hammer rows from the cache, we can use a crafted access pattern~\cite{rowhammerjs} to cause cache conflicts for  hammer rows. Also, we can bypass the cache by accessing uncached memory.

\emph{Second}, the row buffer must be cleared between consecutive hammering DRAM rows. Both double-sided and single-sided hammering explicitly perform alternate access to two or more rows within the same bank to clear the row buffer. 
One-location hammering relies on the memory controller to clear the row buffer.

\emph{Third}, existing rowhammer attacks require that a hammer row be accessible to an attacker in order to gain the privilege escalation or steal the private data, such that a victim row can be compromised by hammering the hammer row. 

\emph{Fourth}, either the hammer row or the victim row must contain sensitive data objects (e.g., page tables) we target. 
If the victim row hosts the data objects, an attacker can either gain the privilege escalation or steal the private data~\cite{seaborn2015exploiting,bhattacharya2016curious}. If the hammer row hosts the data objects, an attacker can steal the private data~\cite{kwong2020rambleed}.

\subsubsection{Rowhammer Attacks}\label{sec:related}
In order to trigger rowhammer bug, frequent and direct memory access is a prerequisite. Thus, we classify rowhammer attacks into three categories based on how they flush or bypass cache. 

\mypara{Instruction-based}
the \texttt{clflush} instruction is commonly used for explicit cache flush~\cite{kim2014flipping, seaborn2015exploiting,gruss2017another,razavi2016flip} ever since Kim et al.~\cite{kim2014flipping} revealed the rowhammer bug. 
Also, Qiao et al.~\cite{qiao2016new} reported that non-temporal store instructions such as \texttt{movnti} and \texttt{movntdqa} can be used to bypass cache and access memory directly.

\mypara{Eviction-based}
alternatively, an attacker can evict a target address by accessing congruent memory addresses which are mapped to the same cache set and same cache slice as the target address~\cite{aweke2016anvil, rowhammerjs, bosman2016dedup,liu2015last,maurice2017hello}. A large enough set of congruent memory addresses is called an eviction set. Our \upperpt also chooses the eviction-based approach to flush Level-1 PTEs from cache. 

\mypara{Uncached Memory-based}
as direct memory access (DMA) memory is uncached, past rowhammer attacks such as Throwhammer~\cite{tatar2018throwhammer} and Nethammer~\cite{lipp2018nethammer} on x86 microarchitecture and Drammer~\cite{van2016drammer} on ARM platform have abused DMA memory for hammering. Note that such attacks hammer target rows that are within an attacker's access permission. 



%% file: overview.tex
\section{\uppertele Overview}\label{sec:overview}
In this section, we first present the threat model and assumptions, and then introduce the formal model of \uppertele, followed by an instance of \uppertele to demonstrate its practicality. 

\subsection{Threat Model and Assumptions}
Our threat model is similar to other rowhammer attacks~\cite{xiao2016one,razavi2016flip,qiao2016new,bosman2016dedup,gruss2016rowhammer,seaborn2015exploiting}. Specifically, 
\begin{itemize}[noitemsep, topsep=2pt, partopsep=0pt,leftmargin=0.4cm]
    \item The kernel is considered to be secure against software-only attacks. In other words, our attack does not rely on any software vulnerabilities.  
   \item An adversary controls an unprivileged user process that has no privileges such as accessing  {\tt pagemap} that has the mapping between a virtual address and a physical address. 
   \item An attacker has no knowledge about the kernel memory locations that are bit-flippable.
   \item The installed DRAM modules are susceptible to rowhammer-induced bit flips. Pessl et al.~\cite{pessl2016drama} report that mainstream DRAM manufacturers have vulnerable DRAM modules, including  DDR3 and DDR4.
\end{itemize}

\subsection{Formal Modeling of \uppertele}\label{sec:telehammer}

We propose a formal model of \uppertele to characterize its attack paradigm.   

Let $U$ be a set of entities, which can be any component in modern OSes that is able to initiate memory accesses. If $u\in U$, then $u$ could be, for instance, a user process, kernel and etc. A set of memory addresses is denoted by $M$. Each memory address has access permissions assigned to each entity.   
Given a memory address $m \in M$, the permission function $\Pi(m)$ returns a set of entities that can access (e.g., read) $m$. 

Only in this model, memory refers to not only the DRAM memory row but also other types of high-speed memory (e.g., cache, register, DRAM row buffer, etc.). 
Generally, a memory access starts by searching the content in the fastest memory hardware first (e.g., registers). If the search fails, then it goes to other memory hardware such as the slowest DRAM memory row.
The validity function ${\mathtt V}(m)\in \{0, 1\}$ indicates whether $m$ contains valid contents (i.e., ${\mathtt V}(m)=1$) or not (i.e., ${\mathtt V}(m)=0$) to satisfy the search. 
The time function ${\mathtt T_{node}}(u, m)$ returns the latency taken by $u$ to access $m$.

As we are aware of, a memory access from an entity may trigger other entities with subsequent memory accesses to complete a computing task. For instance, when a regular user initiates a memory access, it can trigger the modern processor to access page-table entries. 
Such situation is modeled by the directed memory graph defined below.

\begin{figure}
\centering
\includegraphics[scale=0.55]{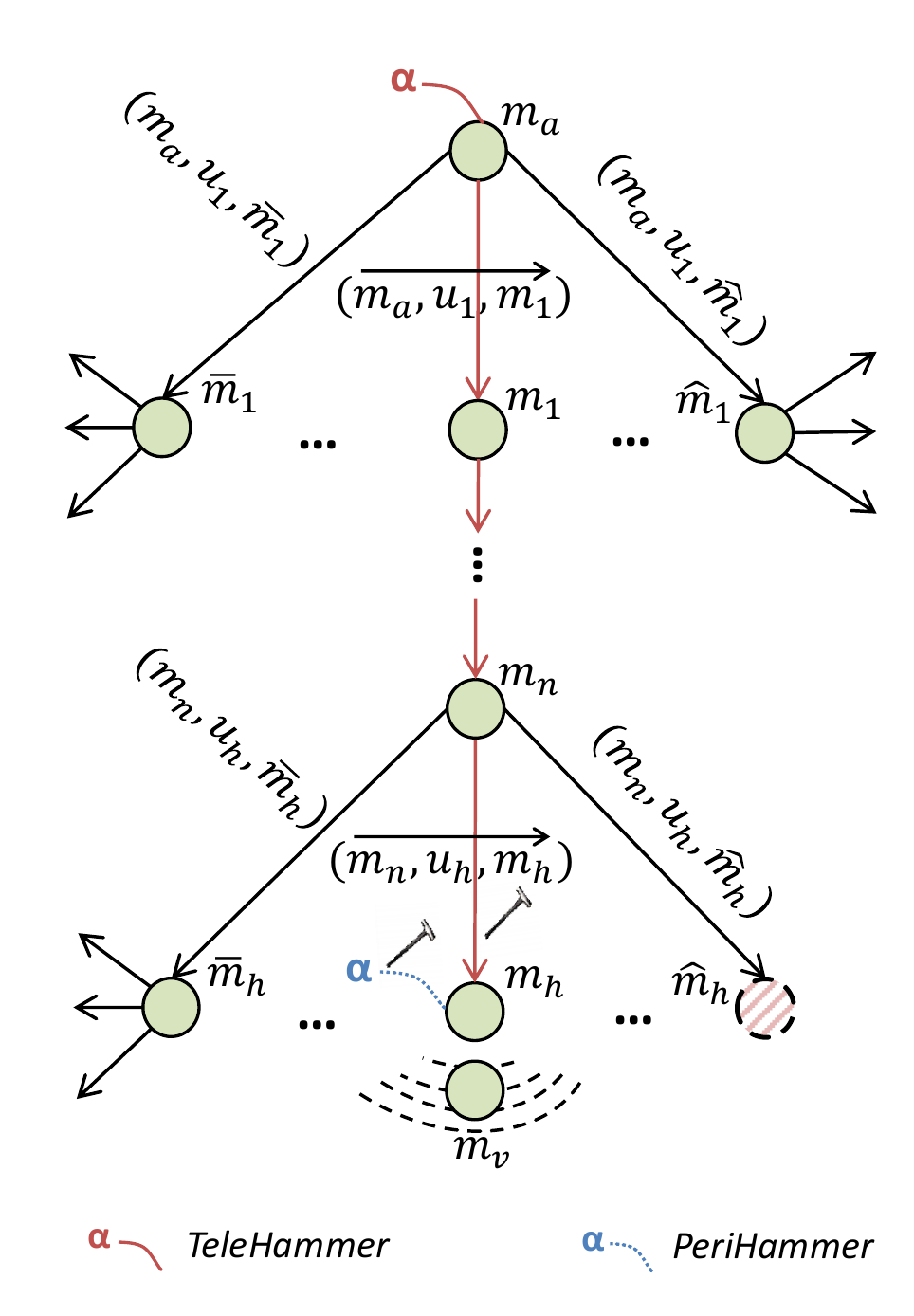}
\caption{Formal Modeling of \uppertele and \upperperi. \uppertele specifies the path from the initial memory access $(\protect\overrightarrow{m_a, u_1, m_1})$ to the last $(\protect\overrightarrow{m_n, u_h, m_h})$ so as to hammer $m_h$ indirectly. When $m_a$ and $m_h$ refer to the same node, i.e., $m_a = m_h$ \uppertele can hammer $m_h$ directly and thus becomes \upperperi. (Node $\hat{m}_h$ has a dashed circle, meaning that ${\mathtt T_{edge}}(m_n, u_h, \hat{m}_h)$ has a lower latency than ${\mathtt T_{edge}}(m_n, u_h, m_h)$, thus $a$ has to set $\hat{m}_h$ to invalid to specify the memory access edge from $m_n$ to $m_h$.)}
\label{fig:hammer}
\end{figure}

\begin{definition}[Directed Memory Graph]
A directed memory graph $G$ (e.g., Figure~\ref{fig:hammer}) is a pair $(M, E)$, where memory addresses in $M$ constitutes the nodes of $G$, and $E$ contains all the directed edges. A directed edge in $E$ is represented by a quintuple such as $(m_a, u_1, m_1)$ in Figure~\ref{fig:hammer}, where $m_a$ and $m_1 \in M$, $u_1 \in U$, respectively. 
An edge $(m_a, u_1, m_1)\in E$ has the following semantics:
\begin{itemize}[noitemsep, topsep=2pt, partopsep=0pt,leftmargin=0.4cm]
    \item $u_1 \in \Pi(m_1)$ and,
    \item an access to $m_a$ can \emph{potentially} trigger $u_1$ to access $m_1$ within time ${\mathtt T_{edge}}(m_a, u_1, m_1)$.
\end{itemize}
\end{definition}
Note that the time ${\mathtt T_{edge}}(m_a, u_1, m_1)$ is decided by $m_a$ triggering $u_1$ and then $u_1$ accessing $m_1$. As such, 
${\mathtt T_{edge}}(m_a, u_1, m_1)$ should be greater than ${\mathtt T_{node}}(u_1, m_1)$ given the time taken by the trigger. 
Since memory addresses in this model have different memory types, there exist other edges starting from $m_a$ such as $(m_a, u_1, \hat{m}_1)$.
Which edge to access at runtime is highly dependent on the time taken by the edge.
Intuitively, the edge that has a shorter time would have a higher chance to be selected.
Take Level-1 cache as an example, it is shared between all the cores of the processor and partitioned into multiple slices (one for each core). Each core will choose to access its own slice rather than others since the time to access its own slice is faster.  

To exploit the rowhammer bug, an attacker must hammer a node (e.g., $m_h$ in Figure~\ref{fig:hammer}) that is located in the DRAM row, rather than other nodes (e.g., $\hat{m}_h$ in cache). As such, the attacker is supposed to select the edge $(m_n, u_h, m_h)$ at runtime and we call such edge a \emph{memory access edge}.



\begin{definition}[Memory Access Edge]
An edge $(m_n, u_h, m_h)$ with ${\mathtt V}(m_h)=1$ is defined as a \emph{memory access edge}, denoted by 
$(\overrightarrow{m_n, u_h, m_h})$ if $\forall (m_n, u_h, \hat{m}_h) \in E$ ($\hat{m}_h \neq m_h$) satisfies the following requirements:
\begin{itemize}[noitemsep, topsep=2pt, partopsep=0pt,leftmargin=0.4cm]
    \item ${\mathtt V}(\hat{m}_h)=0$ or,
    \item ${\mathtt T_{edge}}(m_n, u_h, \hat{m}_h) > { \mathtt T_{edge}}(m_n, u_h, m_h)$.
\end{itemize}
\end{definition}\label{def:accessedge}
If ${\mathtt V}(\hat{m}_h)=0$, such nodes do not contain valid content. Thus, their edges will not be selected at runtime. 
Or if the edges (e.g., $(m_n, u_h, \hat{m}_h)$) take a longer time, then such edges will not be taken, either. 
As runtime, specifying the memory access to $m_h$ can be done by setting ${\mathtt V}(\hat{m}_h)=0$ if ${\mathtt T_{edge}}(m_n, u_h, \hat{m}_h) < { \mathtt T_{edge}}(m_n, u_h, m_h)$. We use a function ${\mathtt T_{set}}(m_h)$ to denote the time cost of specifying the memory access to $m_h$.
For instance, $\hat{m}_h$ and $m_h$ are within the cache and DRAM row, respectively and each holds the same valid data. 
If $u_h$ wants to access $m_h$, $u_h$ could either invoke an instruction (e.g., \emph{clflush}) or leverage previous cache eviction approaches~\cite{aweke2016anvil, rowhammerjs} to flush $\hat{m}_h$, that is, set ${\mathtt V}(\hat{m}_h)$ to $0$. 

If an access to another node such as $m_a$ shown in Figure~\ref{fig:hammer} triggers a chain of memory accesses at runtime, a \emph{communication path} is built up. Formally, the communication path is defined below.




\begin{definition}[Communication Path]
As shown in Figure~\ref{fig:hammer}, $m_1$ and $m_n\in M$ ($n > 1$).
A communication path ${\mathtt P}(m_1,m_n)$ is a sequence of memory access edges ($\vec{e_i}$, $i \in [1,2,...,n-1]$) for which there is a sequence of distinct nodes ($m_i$, $i \in [1,2,...,n]$) such that $\vec{e_i} = (\overrightarrow{m_i, u_{i+1}, m_{i+1}})$ for $i \in [1,2,...,n-1]$.
\end{definition}
Given the path ${\mathtt P}(m_1, m_n)$, we use ${\mathtt last}({\mathtt P}(m_1, m_n))$ to denote 
the last memory access edge in the path. Let $\vec{e}={\mathtt last}({\mathtt P}(m_1, m_n))$. Then, ${\mathtt P}(m_1, m_n)|_{\vec{e}}$ means a subpath of ${\mathtt P}(m_1, m_n)$ excluding $\vec{e}$; that is, the concatenation of ${\mathtt P}(m_1, m_n)|_{\vec{e}}$ and $\vec{e}$ is ${\mathtt P}(m_{n-1}, m_n)$. 
For path ${\mathtt P}(m_1, m_n)$, its time latency is denoted by ${\mathtt T_{p}}({\mathtt P}(m_1, m_n))$.

\begin{definition}[Communication Latency]
Let ${\mathtt P}(m_a, m_h)$ be a communication path, ${\mathtt last}({\mathtt P}(m_a, m_h))=\vec{e}$, and $\vec{e}=(\overrightarrow{m_n, u_h, m_h})$. Then, ${\mathtt T_{p}}({\mathtt P}(m_a, m_h))$ is defined as:
\begin{itemize}[noitemsep, topsep=2pt, partopsep=0pt,leftmargin=0.3cm]
    \item ${\mathtt T_{p}}({\mathtt P}(m_a, m_h))= {\mathtt T_{edge}}(m_n, u_h, m_h)$, if ${\mathtt P}(m_a, m_h) = \vec{e}$,  otherwise,
    \item ${\mathtt T_{p}}({\mathtt P}(m_a, m_h))={\mathtt T_{p}}({\mathtt P}(m_a, m_h)|_{\vec{e}})+{\mathtt T_{edge}}(m_n, u_h, m_h)$.
\end{itemize}
\end{definition}
Note that when ${\mathtt P}(m_a, m_h) = \vec{e}$, then $m_a$ and $m_n$ are the same node, i.e., $m_a = m_n$ such that ${\mathtt P}(m_a, m_h)$ has only one memory access edge.


When a hammer row is being hammered, the rowhammer bug can badly affect either the hammer row itself~\cite{kwong2020rambleed}, or a victim row that is at least one-row away (within the same DRAM bank) from the hammer row~\cite{kim2014flipping, xiao2016one}. 
We use $R_{max}$ to indicate the maximum row distance between a hammer row and a victim row, since some defenses rely on empirically-determined $R_{max}$ for their effectiveness. For example,
ZebRAM~\cite{konoth2018zebram} empirically reported that $R_{max}$ is 1 on its test DRAM modules. However, our experiments in Lenovo X230 show that $R_{max}$ can be 2 (see Section~\ref{sec:dis}). 

Let $m_h$ being the hammered node in DRAM and $m_v$ being another affected DRAM node, then $m_v$ resides either in the same row as $m_h$ or a victim row. We use a row-index function ${\tt Row}(m_h)$ to return the row index of $m_h$ if $m_h$ is within a row, or $-1$ otherwise. As such, the row distance between ${\tt Row}(m_h)$ and ${\tt Row}(m_v)$ should be no greater than $R_{max}$.
To make the rowhammer bug exploitable, $m_v$ contains sensitive information (e.g., a page table or a cryptographic key), making sensitivity function ${\mathtt S}(m_v)$ return $1$, otherwise $0$. 

As a minimum hammering frequency is required to hammer $m_h$ and trigger the rowhammer bug,  we use ${\mathtt T_{max}}$ to represent a required maximum time latency per hammering. ${\mathtt T_{max}}$ is highly dependent on a DRAM module and the hammering technique.
For example, we perform double-sided rowhammer test in the Lenovo X230 in Section~\ref{sec:timecosts} and the maximum time cost per double-sided hammering (i.e., ${\mathtt T_{max}}$) should be less than 1500 cycles to trigger rowhammer bit flips. 
 
To this end, the following defines the necessary conditions for a \uppertele based exploit.
\begin{definition}[\uppertele]\label{def:telehammer}
Let $G$ be the directed memory graph of a computing task being conducted by an attack process $a$, exemplified in Figure~\ref{fig:hammer}, where $m_a$,
$m_h$ and $m_v$ ($a \not\in \Pi(m_v)$) represent an attack address, a hammer address and a victim address, respectively. 
$a$ can launch a \uppertele-based exploit, if conditions below are satisfied:     

\begin{itemize}[noitemsep, topsep=2pt, partopsep=0pt,leftmargin=0.4cm]
    \item ${\tt Row}(m_h) \neq -1, {\tt Row}(m_v) \neq -1$, ${\mathtt S}(m_v) = 1$,
    \item $\mid{\tt Row}(m_h) - {\tt Row}(m_v)\mid  \leq R_{max}$, 
    \item $\exists {\mathtt P}(m_a, m_h)$ in $G$, 
    \item ${\mathtt T_{set}}({m_h}) + {\mathtt T_{node}}(a, m_a) + {\mathtt T_{p}}({\mathtt P}(m_a, m_h)) + {\mathtt T_\delta} \leq {\mathtt T_{max}}$
\end{itemize}
\end{definition}
%
The last condition specifies the time requirement for \uppertele. As shown in Figure~\ref{fig:hammer}, modern hardware expects to take the fastest path to handle the computing task for the attack process $a$, i.e., ${\mathtt P}(m_a, \hat{m}_h)$. 
However, the path must be changed to ${\mathtt P}(m_a, m_h)$ so as to hammer $m_h$ by using $u_h$. As $u_h$ accessing nodes such as $\hat{m}_h$ takes a shorter time, $a$ must set such nodes invalid to specify $m_h$ for $u_h$ to access.
The time taken by the specifying is ${\mathtt T_{set}}(m_h)$. 

Also, $a$ considers the time by accessing $m_a$ (i.e., ${\mathtt T_{node}}(a, m_a)$), the time by walking through the communication path ${\mathtt P}(m_a, u_h, m_h)$ (i.e., ${\mathtt T_{p}}({\mathtt P}(m_a, m_h))$) and the time  $a$ has to wait to perform next hammering (i.e., ${\mathtt T_\delta}$). 

Note that the value of ${\mathtt T_\delta}$ is dependent on whether $m_h$ is last node that a computing task needs to access. As $m_h$ is in Figure~\ref{fig:hammer}, ${\mathtt T_\delta}$ is then negligible. Otherwise, $a$ waits for ${\mathtt T_\delta}$, in which period the computing task reaches last node from $m_h$.
Take the task of one system call handler as an example, if $a$ targets $m_h$ that hosts one entry of system-call table for hammering, $a$ needs to wait for the system call handler to complete its routine after $m_h$ is accessed. 

When $m_a$ and $m_h$ refer to the same memory address, i.e., $m_a = m_h$, then \uppertele has an access to $m_h$ and actually becomes \upperperi. As such,  we can also formally define \upperperi below based on the above formal model.

\begin{definition}[\upperperi] \label{def:perihammer}
\upperperi would succeed if the following conditions are met:

\begin{itemize}[noitemsep, topsep=2pt, partopsep=0pt,leftmargin=0.4cm]
    \item ${\tt Row}(m_h) \neq -1, {\tt Row}(m_v) \neq -1$, ${\mathtt S}(m_v) = 1$,
    \item $\mid{\tt Row}(m_h) - {\tt Row}(m_v)\mid \leq R_{max}$, 
    \item $a \in \Pi(m_h)$,
    \item ${\mathtt T_{set}}({m_h}) + {\mathtt T_{node}}(a, m_h) + {\mathtt T_{\delta}} \leq {\mathtt T_{max}}$
\end{itemize}
\end{definition}
Clearly, the last condition removes the latency caused by the path ${\mathtt P}(m_a, m_h)$, making it faster to hammer once. Besides, it is much easier for $a$ to specify the access to $m_h$ other than other nodes and spend much less time 
compared to $a$ in \uppertele as discussed in Definition~\ref{def:accessedge}. ${\mathtt T_{\delta}}$ is neglectable since $m_h$ is the only accessed node and $a$ does not have to wait for next hammering. 

\mypara{A comparison of \uppertele and \upperperi}
as shown in Figure~\ref{fig:hammer}, \uppertele is \emph{effective} against the rowhammer defenses where $m_h$ is located in a physically isolated DRAM partition, since it requires no access to $m_h$.

On top of that, \uppertele is \emph{stealthy} and hard to be traced by dynamic analysis at runtime, since it has a complicated communication path and hammer $m_h$ by using $u_h$. In contrast, an attacker via \upperperi hammers $m_h$ directly by herself.

Besides, mitigating \uppertele is \emph{challenging} due to abundant communication path candidates. \uppertele can identify as many paths as possible by leveraging built-in features of modern hardware and/or software. Thus eliminating the communication path we have identified in the following sections essentially cannot defend against \uppertele. 

Clearly, \uppertele is slower than \upperperi by comparing their time condition in their respective definition, indicating that \upperperi is potentially faster in inducing bit flips.









To demonstrate the practicality of \uppertele, we have created an instance of it, called \upperpt. Besides, we discuss in detail about other possible instances in Section~\ref{sec:dis}.

\subsection{\upperpt: page-table based \uppertele}\label{sec:pthammer}
\upperpt is page-table based \uppertele. It allows an unprivileged attacker to cross privilege boundary and hammer page tables, resulting in bit flips in other page tables. 
In the following, we discuss how \upperpt satisfies the formal conditions in Definition~\ref{def:telehammer}.

\subsubsection{Satisfy Formal Definition~\ref{def:telehammer}}
\emph{First,} page tables are critical in memory isolation and are inaccessible to an unprivileged attacker. If the attacker can compromise a memory address hosting page tables by the rowhammer effect, then it satisfies the first condition; here, the address refers to $m_v$ and ${\mathtt S}(m_v) = 1$.  

\emph{Second,} page tables are common and can be widely distributed in modern OS kernels. Thus, both $m_h$ and $m_v$ can be kernel addresses hosting page tables, that is, hammering page tables of $m_h$ will flip bits in page tables of $m_v$. To this end, we can leverage previous works such as memory spray~\cite{cheng2018still} and memory ambush ~\cite{seaborn2015exploiting} to force the kernel to create a large number of page-table pages, with the hope that some page tables are placed into hammer addresses like $m_h$ while some are within victim addresses like $m_v$.
As such, we can create numerous pairs of such hammer addresses and victim addresses so that they become highly likely to induce exploitable bit flips in page tables. 
Note that the rowhammer defense (i.e., CTA~\cite{wu2018CAT}) allocates all page tables from a reserved memory partition and this will greatly increase the number of pairs compared to page-table allocation from the whole system memory.

\begin{figure}
\centering
\includegraphics[scale=0.45]{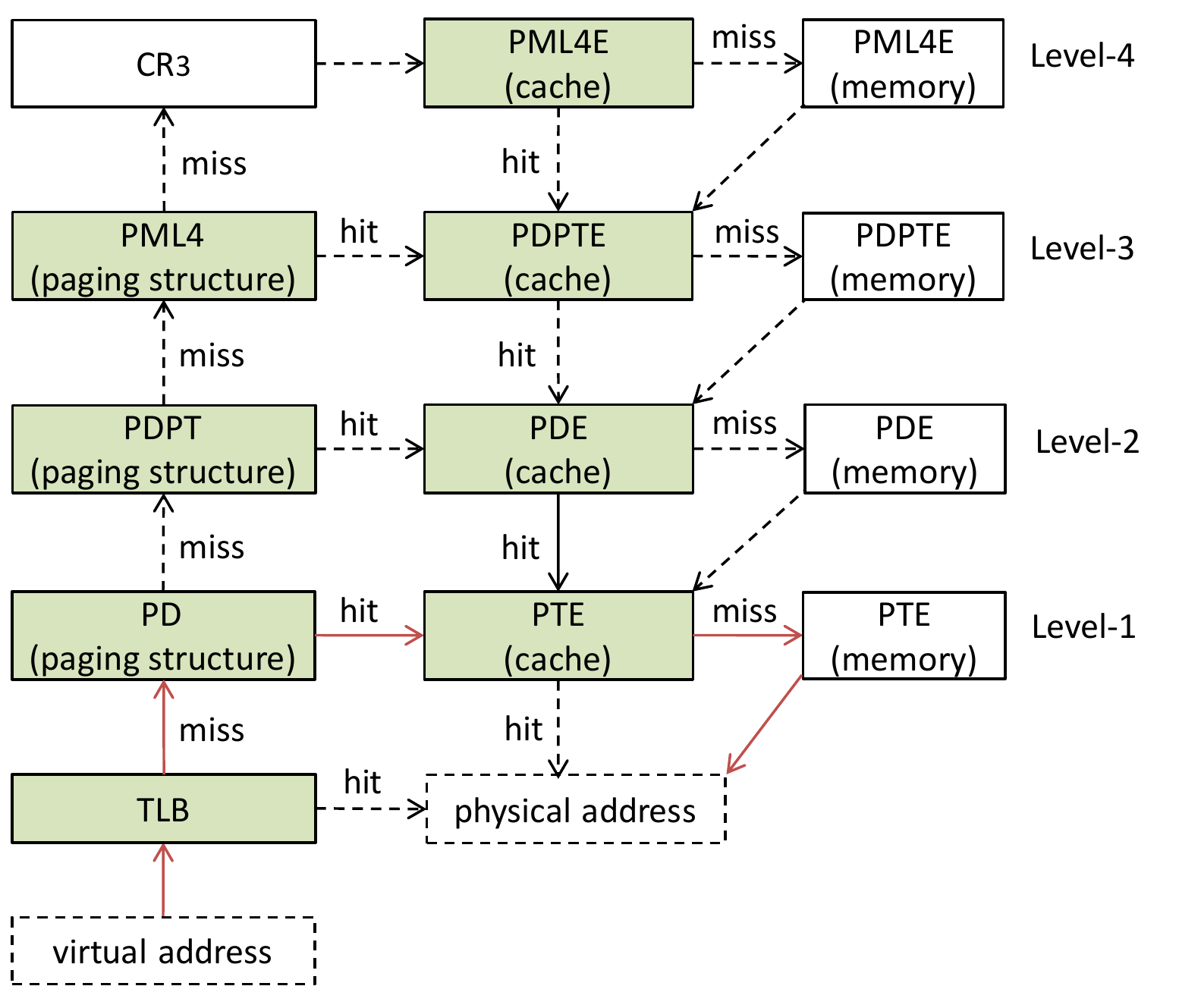}
\caption{Address Translation. A solid line with an arrow indicates the fastest communication path that \upperpt identifies to hammer a Level-1 page-table entry (PTE). When specifying the path, \upperpt only flushes TLB and cache while retains all-level paging structure effective. Note that PML4E, PDPTE, PDE are other three-level PTE, respectively.} 
\label{fig:pt_walk}
\end{figure}

\emph{Third,} there exists a \emph{communication path} (see Definition~\ref{def:telehammer}) that allows an attacker to indirectly access page tables. To this end, we observe that a least privileged memory triggers an address translation where the processor can access page tables from memory.
When a user allocates a virtual memory page by \texttt{malloc} and then accesses the page for the first time, an address-translation process occurs. Within the process, the processor performs multi-level page-table walk, populates corresponding 
page-table entries (PTEs) and allocates a physical memory page for the user. 
To facilitate subsequent memory access as shown in Figure~\ref{fig:pt_walk},
Translation Look-aside Buffer (TLB) stores a complete address mapping from a virtual address to a physical address. Paging structure caches a partial address mapping of different page-table levels~\cite{barr2010translation}. For instance, the paging structure of Level-2 PD translates a virtual address to a physical address of Level-1 PT. With bits 12$\sim$20 from the virtual address~\cite{intelOp}, a corresponding physical address of a Level-1 PTE (L1PTE) can be obtained. 
CPU cache copies the accessed four-level PTEs from memory. 
By doing so, the processor will search these hardware structures in the order of priority to get a matching physical address.
If TLB, paging structure and cache are all effectively flushed, the processor then has to access the four-level PTEs from memory.
As such, an access to an address $m_a$ by $a$ can trigger the processor to access four-level PTEs from memory if the flushing operation is conducted effectively.

\emph{Last,} as $m_a$ can be within cache, the time (${\mathtt T_{node}}(a, m_a)$) to access it is negligible. 
To meet the time condition in the definition, 
the time (${\mathtt T_{p}}({\mathtt P}(m_a, m_h))$) to walk through the above identified path, the time (${\mathtt T_{set}}(m_h)$) to specify the path, and the time (${\mathtt T_\delta}$) to wait for next hammering must all be as low as possible. 

\mypara{Optimize ${\mathtt T_{p}}({\mathtt P}(m_a, m_h))$ and ${\mathtt T_\delta}$}
we optimize the identified communication path by making $m_h$ host one L1PTE rather than other-level PTE, shown as a solid line with an arrow in Figure~\ref{fig:pt_walk}. The path is optimized for  following reasons.
\begin{itemize}[noitemsep, topsep=2pt, partopsep=0pt,leftmargin=0.4cm]
    \item flushing paging structure is required when $m_h$ hosts other-level PTE. Directly flushing paging structure requires executing a privileged instruction such as \texttt{invlpg}, while indirect flushing needs to reverse-engineer the mapping between a virtual address and the paging structure index. 
    \item flushing all-level (or other single-level) PTEs from paging structure and cache is intuitively more time consuming compared to flushing the L1PTE, as shown in Figure~\ref{fig:pt_walk}. 
    \item specifying such a path consumes much less memory, making the exploit stealthier. As mentioned above, we need to allocate a lot of page-table pages to flip exploitable bits in page tables. Creating a PT-level page of $512$ entries requires exhausting 2MiB memory. For a higher page-table-level page, its creation requires much more memory. For example,  the PD-level page creation requires 1GiB memory.
    \item compared to other-level PTEs, a bit flip in one L1PTE is easier to become exploitable and gain privilege escalation, since the L1PTE decides the physical page that a user can access as well as the access permission. 
\end{itemize}
As $m_h$ that hosts one L1PTE is the last accessed memory address when translating a virtual address, it means that ${\mathtt T_\delta} \approx 0$.

\mypara{Optimize ${\mathtt T_{set}}(m_h)$}
to specify the path to the L1PTE (shown in red solid line in Figure~\ref{fig:pt_walk}), we only need to flush the address mapping from TLB and the L1PTE from cache with the PD-level paging structure still being effective. 

Intuitively, we can simply invoke \texttt{invlpg} to flush the whole TLB. As for the cache flush, we can perform a page-table walk, get the virtual address of the L1PTE and thus flush its valid content from cache by invoking \texttt{clflush}. 
By doing so, we are able to flush both TLB and cache as quickly as possible. 
However, kernel privilege is required to complete the flushing. Alternatively, we can perform the flushing indirectly by manipulating cache and TLB replacement states. 
As the size of TLB and cache is limited, we can simply create many pages as an eviction buffer and access those pages one by one so as to evict target TLB entry and cache line. Although this approach can effectively flush both TLB and cache, it does not reduce ${\mathtt T_{set}}(m_h)$ to its minimum. 

In a nutshell, the key challenge to minimize the time is to determine two minimum eviction sets so as to flush targeted TLB entry and cache line \emph{effectively} and \emph{efficiently}.

\subsubsection{{Effective and Efficient TLB Flush}}\label{sec:tlbflush}

As Gras et al.~\cite{gras2018translation} have revealed that there exists an explicit mapping between a virtual page number and multi-level TLB set, we simply create an initial eviction set that contains multiple (physical) pages to flush a cached virtual address from TLB. One subset of the pages is congruent and mapped to a same L1 dTLB set while the other is congruent and mapped to a same L2 sTLB set if TLB applies a non-inclusive policy. 

Take one of our test machines, Lenovo T420, as an example, both L1 dTLB and L2 sTLB have a $4$-way set-associative for every TLB set and thus $8$ (physical) pages are enough as an minimum eviction set to evict a target virtual address from TLB. However, when we create such an eviction set and then profile the access latency of a target virtual address, its latency remains unstable. 
To collect finer-grained information on TLB misses induced by the target address, we develop a kernel module that applies Intel Performance Counters (PMCs) to monitor the TLB-miss event (i.e., \texttt{dtlb\_load\_misses.miss\_causes\_a\_walk}). The experimental results show that TLB misses in both levels do not always occur when profiling the target address, meaning that the target address has not been effectively evicted by the eviction set, and thereby rendering the TLB flush ineffective. A possible reason is that the eviction policy on TLB is not true Least Recently Used (LRU). 

\mypara{Decide the Minimal Size for a TLB Eviction Set}
to this end, we propose a working Algorithm~\ref{alg:tlb_flush} that can decide a minimal size without knowing its eviction policy. Note that the minimal size is used to prepare a minimal TLB eviction set in \upperpt while \upperpt itself does not use the algorithm.  
Specifically, line 2 to 8 defines a function $profile\_tlb\_set$ that reports a TLB-miss number ($tlb\_miss\_num$) induced by accessing $target\_addr$. Specifically,
the function argument (i.e., $set$) is write-accessed (line 4-6) to flush the cached $target\_addr$ in TLB (line 3) and then $tlb\_miss\_num$ of write-accessing $target\_addr$ is reported in line 7.
Based on a pre-allocated $buf$, we select those all pages that are indexed to the same TLB set as the $target\_addr$ by leveraging the reverse engineered mapping~\cite{gras2018translation} in line 9-14. 
Note that the $buf$ size is large enough to effectively flush any targeted virtual address and it is decided by the number of TLB entries that serve 4KiB-page translation if $target\_addr$ is allocated from a 4KiB-page list, otherwise, the number of TLB entries that support 2MiB or 1GiB should be involved. 
The selected pages are then populated and added into $init\_set$ as shown in line 10-13. It is necessary to populate the selected pages in order to trigger the address-translation feature and thus TLB will cache address mappings accordingly. 
In line 15, we can gain a threshold for effective TLB flushes.
We then start to tailor the set to its minimum while retain its effectiveness in line 16-23.

\begin{algorithm}[t]
\small
	\caption{Decide a minimal eviction-set size for TLB }\label{alg:tlb_flush}
	\textbf{Initially:} $target\_addr$ is a page-aligned virtual address that needs its cached TLB entry flushed. A buffer ($buf$) is pre-allocated, size of which is decided by available TLB entries. A set ($init\_set$) is initialized to empty. A unique number is assigned to $data\_marker$.\\
	\SetKwProg{Fn}{Function}{}{}
	    \Fn {$profile\_tlb\_set (set)$} {
	        $target\_addr \leftarrow data\_marker$ \\
	        \ForEach {$page  \in set$} {
	            $page[0]  \leftarrow data\_marker$ \\
	        }
	        $tlb\_miss\_num$ is decided by accessing $target\_addr$. \\
	       \KwRet $tlb\_miss\_num$ \\
        }
	   \ForEach {$page \in buf$} {
            \If {$page$ $and$ $target\_addr$ $are$ $in$ $the$ $same$ $set$} {
            $page[0]   \leftarrow data\_marker$ \\
            add $page$ into $init\_set$. \\
        }
      }  
	   $threshold   \leftarrow profile\_tlb\_set(init\_set)$ \\
	    
	   \For{$page \in init\_set$} {
	       take one $page$ out of $init\_set$. \\
	       $temp\_tlb\_miss  \leftarrow profile\_tlb\_set(init\_set)$ \\
	    \If {$temp\_tlb\_miss < threshold$} {
	        put $page$ back into $init\_set$ and break. \\
	    }
	   }
	   \KwRet the size of $init\_set$ \\
\end{algorithm}

\subsubsection{{Effective and Efficient Cache Flush}}\label{sec:cacheflush}
Now we are going to flush a cached Level-1 PTE (L1PTE) that corresponds to a target virtual address. 
Considering that last-level cache (LLC) is inclusive~\cite{intelOp}, we target flushing the L1PTE from LLC such that the L1PTE will also be flushed out from both L1 and L2 caches (we thus use cache and LLC interchangeably in the following section).
In contrast to TLB that is addressed by a virtual page-frame number, LLC is indexed by physical-address bits, the mapping between them has also been reverse engineered~\cite{hund2013practical,Maurice2015,irazoqui2015systematic}. Based on the mapping, we can intuitively create an eviction set consisting of many congruent memory lines (i.e.,cache-line-aligned virtual addresses), which are mapped to the same cache slice and cache set as the L1PTE. 
On top of that, the eviction set can also be minimized in case where the eviction policy of LLC is not publicly documented. 

\mypara{Decide the Minimal Size for an LLC Eviction Set}
we extend the aforementioned kernel module to count the event of LLC misses (i.e., \texttt{longest\_lat\_cache.miss}) and have a similar algorithm to Algorithm~\ref{alg:tlb_flush} to decide the minimal size for an LLC eviction set, namely, prepare a large enough eviction set congruent as a target virtual address and gain a threshold of LLC-miss number induced by accessing the target address, remove memory lines randomly from the set one by one and verify whether currently induced LLC-miss number is less than the threshold. If yes, a minimal size can be determined. Also, this algorithm is performed in an offline phase long before \upperpt is launched. 

Although the size of eviction-set is determined ahead of time, \upperpt in our threat model cannot know the mapping between a virtual and a physical address, making it challenging to prepare an eviction set for any target virtual address during its execution. Also, \upperpt cannot obtain the L1PTE's physical address, and thus it is difficult to learn the L1PTE's exact location (e.g., cache set and cache slice) in LLC.
To address the above two problems, \upperpt at the  beginning prepares a complete pool of eviction sets, which can be used to flush any target data object including the L1PTE. It then selects an eviction set from the pool to evict a target L1PTE without knowing the L1PTE's cache location. Note that preparing the eviction pool is a one-off cost and \upperpt only need to repeatedly select eviction sets from the pool when hammering L1PTEs.

\mypara{Prepare a Complete Pool of LLC Eviction Sets}
the pool has a large enough number of eviction sets and each can be used to flush a memory line from a specific cache set within a cache slice in LLC. The size of each eviction set is the pre-determined minimum size. We implement the preparation based on previous works~\cite{liu2015last,genkin2018drive}. Both works  rely on the observation that a program can determine whether a target line is cached or not by profiling its access latency. If a candidate set of memory lines is its eviction set, then the target line's access latency is above a time threshold after iterating all the memory lines within the candidate set. 

Specifically, if a target system enables \emph{superpage}, a virtual address and its corresponding physical address have the same least significant 21 bits, indicating that if we know a virtual address from a pre-allocated super page, then its physical address bit 0$\sim$20 is leaked and thus we know the cache set index that the virtual address maps to (see Section~\ref{sec:cache}). The only unsolved is the cache slice index. Based on a past algorithm~\cite{liu2015last}, we allocate a large enough memory buffer (e.g., twice the size of LLC), select memory lines from the buffer that have the same cache-set index and group them into different eviction sets, each for one cache slice. 

If \emph{superpage} is disabled, then only the least significant 12 bits (i.e., 4KiB-page offset) is shared between virtual and physical addresses and consequently we know a partial cache-set index (i.e., bits 6$\sim$11). As such, we utilize another previous work~\cite{genkin2018drive} to group potentially congruent memory lines into a complete pool of individual eviction sets. Compared to the above grouping operation, this grouping process is much slower, since there are  many more memory lines sharing the same partial cache-set bits rather than complete bits.

\begin{algorithm}[t]
	\caption{Select a minimal LLC eviction set }\label{alg:cache_flush}
	\small
	\SetAlgoLined
	\textbf{Initially:} a virtual page-aligned address ($target\_addr$) is allocated and needs its L1PTE cache-line flushed. A complete pool of individual eviction sets ($eviction\_sets$). $l1pte\_offset$ is decided by $target\_addr$. $max\_latency$ is initialized to $0$ and indicates the maximum latency induced by accessing $target\_addr$. $max\_set$ represents the eviction set used for the L1PTE cache flush. \\
		\SetKwProg{Fn}{Function}{}{}
	    \Fn {$profile\_evict\_set (set, target)$} {
	        \ForEach {$memory\_line \in set$} {
	            read-access $memory\_line$.
	        }
	        flush a target TLB entry. \\
	        $latency$ is decided by accessing $target$. \\
	       \KwRet $latency$ \\
        }
	\ForEach {$set \in eviction\_sets$} {
	      obtain $page\_offset$ from first memory line in $set$.\\
	      \If {$page\_offset == l1pte\_offset$} {
	         $latency \leftarrow profile\_evict\_set(set, target\_addr)$. \\
	         \If {$max\_latency < latency$} {
	            $max\_latency = latency$. \\
	            $max\_set = set$.\\
	         }
	      }
	     
	}
	\KwRet $max\_set$ \\
\end{algorithm}

\begin{table*}
\centering
\begin{tabular}{cccccc}
\hline
\multirow{2}{*}{\textbf{Machine Model}} & \multirow{2}{*}{\textbf{Architecture}} & \multicolumn{3}{c}{\textbf{CPU}} & 
\multirow{2}{*}{\textbf{DRAM}} \\ \cline{3-5}
 &  &  {Version} & {TLB (Wayness)} & {LLC (Wayness, Size)} &  \\ \hline
\multirow{2}{*}{Lenovo T420} & \multirow{2}{*}{Sandy Bridge} & \multirow{2}{*}{i5-2540M} & \multirow{2}{*}{4-way L1dTLB, 4-way L2sTLB} & \multirow{2}{*}{12-way, 3MiB} & \multirow{2}{*}{Samsung DDR3 8GiB} \\
 &  &  &  &  \\ \hline

\multirow{2}{*}{Lenovo X230} & 
\multirow{2}{*}{Ivy Bridge} & \multirow{2}{*}{i5-3230M} & \multirow{2}{*}{4-way L1dTLB, 4-way L2sTLB} & \multirow{2}{*}{12-way, 3MiB} & \multirow{2}{*}{Samsung DDR3 8GiB} \\
 &  &  &  &  \\ \hline
\multirow{2}{*}{Dell E6420} & 
\multirow{2}{*}{Sandy Bridge} & 
\multirow{2}{*}{i7-2640M} & \multirow{2}{*}{4-way L1dTLB, 4-way L2sTLB}  &
\multirow{2}{*}{16-way, 4MiB} & \multirow{2}{*}{Samsung DDR3 8GiB} \\
 &  &  &  &  \\ \hline
\end{tabular}
\caption{System Configurations.}
\label{tab:config}
\end{table*}

\mypara{Select a Target LLC Eviction Set}
based on the pool preparation, we develop an Algorithm~\ref{alg:cache_flush} to select an eviction set from the pool and evict a L1PTE corresponding to a target address.

In line 9, we enumerate all the eviction sets in the pool and collect those sets that have the same page offset as the L1PTE in line 11. 
This collect policy is based on an interesting property of the cache. Oren et al.~\cite{oren2015spy} report that 
if there are two different physical memory pages that their first memory lines are mapped to the same cache set of LLC, then the rest memory lines of the two pages also share (different) cache sets. This means if we request many (physical) memory lines that have the same page offset as the L1PTE and access each memory line, then we can flush the L1PTE from LLC.

After the selection, line 12-16 will select the target eviction set from the collected ones.
In line 12, we profile every selected eviction set through a predefined function from line 2-8. Within this function, we perform read access to each memory line of one eviction set, which will implicitly flush the L1PTE from LLC if the eviction set is congruent with the L1PTE, and then flush the target TLB entry related to $target\_addr$ to make sure the subsequent address translation will access the L1PTE. At last, we measure the latency induced by accessing $target\_addr$. Based on this function, we can find the targeted eviction set that causes the maximum latency in line 13-16, as fetching the L1PTE from DRAM is time-consuming when accessing $target\_addr$ triggers the address translation in line 7. 
Give that LLC is shared between page-table entries and user data, we must carefully set $target\_addr$ to page-aligned (normally 4KiB-aligned) but not \emph{superpage}-aligned (normally 2MiB-aligned), that is, its page offset is $0$ and different from $l1pte\_offset$, which is the page offset of L1PTE. As such, they are placed into different cache sets and the selected eviction set is ensured to flush the target L1PTE rather than $target\_addr$. 

%% file: eva.tex
\section{Evaluation}\label{sec:eva}
In this section, we test \upperpt on three different machines running a Ubuntu system shown in Table~\ref{tab:config} and each Ubuntu system by default disables the \emph{superpage} feature. 
No matter whether \emph{superpage} is enabled or not, we can observe the first cross-boundary bit flip on each test machine. As a case, we then leverage \upperpt to compromise the state-of-the-art rowhammer defenses on Lenovo T420 with the default system setting. 

\begin{figure*}[ht]
	\centering
	\begin{subfigure}[t]{\columnwidth}
		\centering
		\includegraphics[scale=0.5]{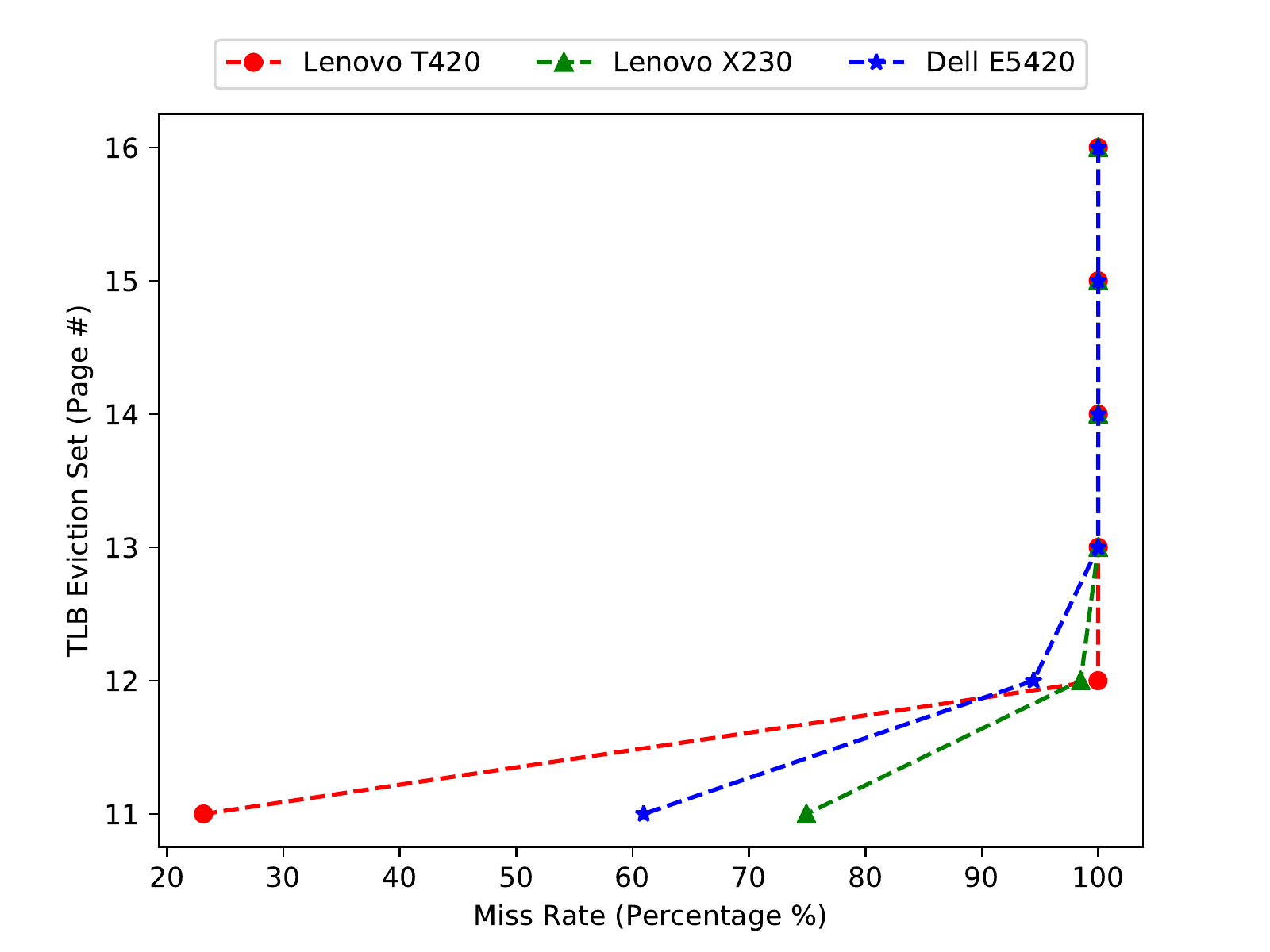}
		\caption{The TLB miss rate on each machine remains relatively stable (no more than 95\%) when the TLB eviction set reduces from $16$ to $12$ and then decreases dramatically when the set is reduced to $11$.}
		\label{fig:tlbflush}
	\end{subfigure}
	\hfill
	\begin{subfigure}[t]{\columnwidth}
		\centering
		\includegraphics[scale=0.5]{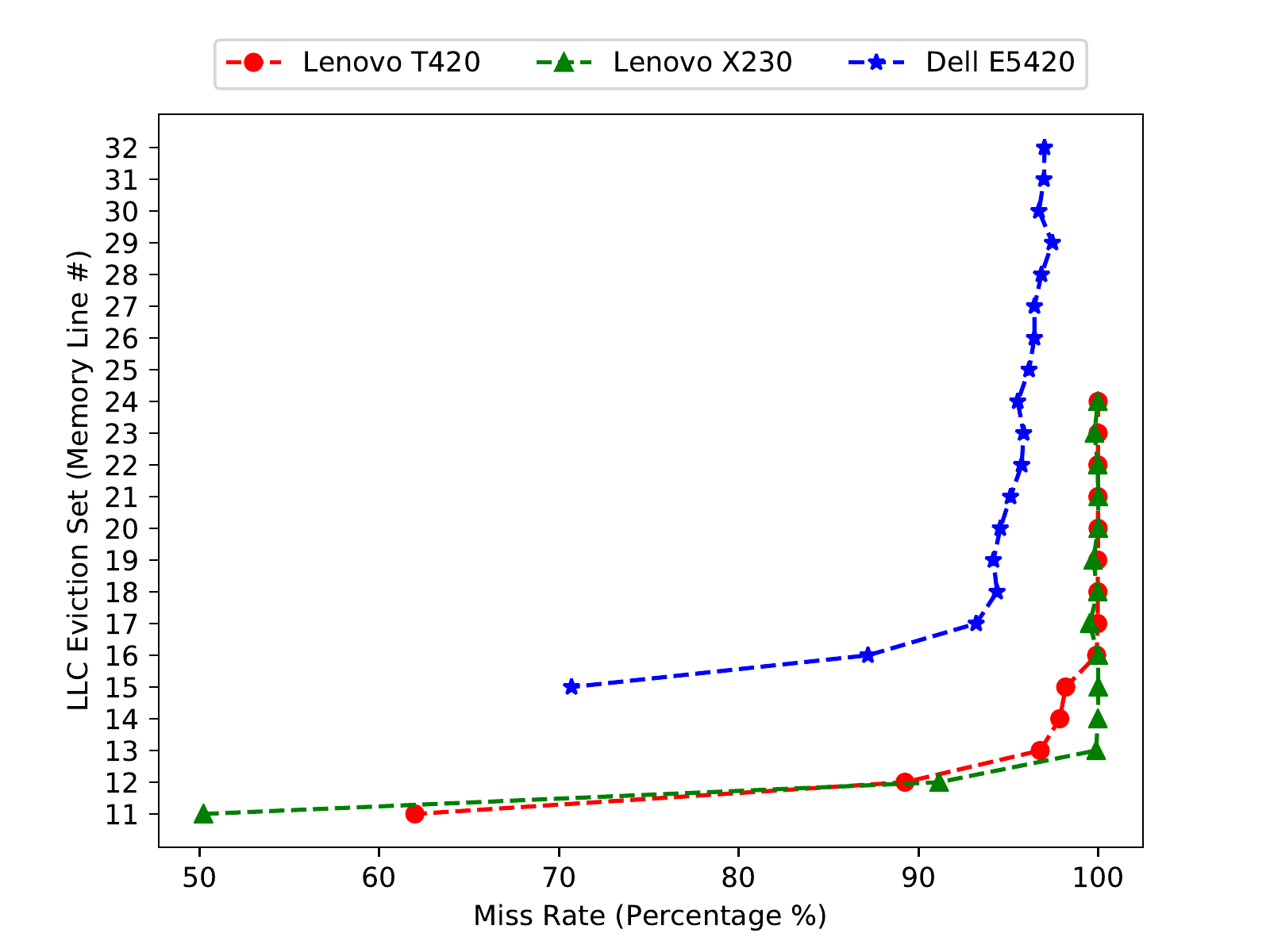}
		\caption{The LLC miss rate on the two Lenovo machines stays greater than 95\% when their eviction set is no more than 13 while Dell machine still has a high miss rate of 94\% when its eviction set is 17.}
		\label{fig:cacheflush}
	\end{subfigure}%
	\caption{TLB/LLC miss rate with regard to the size of TLB/LLC eviction set.}
	\label{fig:flush}
\end{figure*}


\begin{table*}
\centering
\begin{tabular}{ccccccccc}
\hline
\multirow{2}{*}{\textbf{Machine Model}} & \multirow{2}{*}{\textbf{System Setting}} &
\multicolumn{2}{c}{\textbf{One-off Eviction Pool Preparation}} & \multicolumn{2}{c}{\textbf{{Eviction Set Selection}}} & \multirow{1}{*}{\textbf{Hammering Cost}}  \\
& & TLB & LLC & TLB & LLC & \multirow{1}{*}{\textbf{Until First Bit Flip}}  \\ 
\hline
\multirow{2}{*}{Lenovo T420} & \multirow{1}{*}{\emph{superpage}} & 11\emph{millisec} & 0.3\emph{min} & 1\emph{microsec} & 285\emph{millisec} & 10\emph{min}  \\
& \multirow{1}{*}{\emph{regularpage}} & 11\emph{millisec} & 18\emph{min} & 1\emph{microsec} & 283\emph{millisec} & 10\emph{min}  \\ 
\hline
\multirow{2}{*}{Lenovo X230} & \multirow{1}{*}{\emph{superpage}} &  7\emph{millisec} & 0.3\emph{min} & 1\emph{microsec} & 282\emph{millisec} & 15\emph{min} \\
& \multirow{1}{*}{\emph{regularpage}} & 7\emph{millisec} & 19\emph{min} & 1\emph{microsec} & 288\emph{millisec} & 15\emph{min} \\
\hline
\multirow{2}{*}{Dell E6420} & \multirow{1}{*}{\emph{superpage}} & 7\emph{millisec} & 0.3\emph{min} & 1\emph{microsec} & 258\emph{millisec} & 14\emph{min} \\
& \multirow{1}{*}{\emph{regularpage}} & 7\emph{millisec} & 38\emph{min} & 1\emph{microsec} & 270\emph{millisec} & 12\emph{min} \\
 \hline
\end{tabular}
\caption{Averaged time costs for conducting \upperpt on each machine for 5 runs. The first bit flip can be observed within 15 minutes of double-sided hammering.
The pool preparation for either TLB or LLC is a one-off cost and performed only once at the beginning of \upperpt such that respective TLB and LLC eviction set are selected each time right before hammering.}
\label{tab:timecosts}
\end{table*}

\subsection{\upperpt}\label{sec:pthammereva}
We first decide the minimal eviction-set size to effectively and efficiently flush TLB and last-level cache (LLC) at an offline stage. Based on the minimal size, we can prepare a minimal TLB or LLC eviction set from a complete pool of TLB or LLC eviction sets.

\subsubsection{Respective Minimal Eviction-Set Size}
Based on the Algorithm~\ref{alg:tlb_flush} in Section~\ref{sec:tlbflush}, we first obtain an initial eviction set where its page number is twice the number of both L1dTLB and L2sTLB wayness and each page in the set is mapped to the same L1dTLB set or L2sTLB set as a target page-aligned virtual address. We then remove one page from the set each time to check a TLB miss rate of the target virtual address, as shown in Figure~\ref{fig:tlbflush}.
As all the test machines have the same wayness of L1dTLB and L2sTLB, the TLB miss rate on each machine initially remains relatively stable when the eviction-set size drops down one by one until 12, and thereafter decreases dramatically. We choose 12 as the minimal TLB eviction-set size.

For LLC, each Lenovo machine has 12-wayness of LLC and thus each initial eviction set is set to 24 memory lines that map to the same LLC set as a target virtual address. For the Dell machine, it has 16-wayness and thus its initial set size is 32. 
Similar to TLB, memory lines in the eviction set are also removed one by one and the LLC miss rate for each removal is shown in Figure~\ref{fig:cacheflush}. 
Clearly, the LLC miss rate on the Lenovo machines stays stable (more than 95\%) until the set size of 13, but decreases gradually below 90\% after 12.
For the Dell machine, its miss rate drops below 94\% when its set size is less than 17.
As such, we choose 13 as the minimal eviction-set size for the Lenovo machines and 17 for the Dell machine.

\subsubsection{One-off Respective Eviction Pool Preparation}
%
For TLB, we allocate a complete pool of 4KiB pages and its page number is 8 times as many as the number of both L1dTLB and L2sTLB entries that translate a 4KiB-page. 
As can be see from Table~\ref{tab:timecosts}, TLB pool preparation in each setting on each test machine is pretty fast (within milliseconds). 

For LLC, we prepare a complete pool of either 2MiB pages (i.e., \emph{superpage} enabled) or 4KiB pages (i.e., \emph{regularpage} by default) and its size in both settings are twice the size of LLC. As the cache-set bits in the \emph{superpage} setting are all known, the eviction pool preparation is much faster compared to that in the \emph{regularpage} setting. 
Particularly, in the \emph{regularpage} setting, the pool preparation in the Dell machine takes nearly 20 minutes longer than that in the two Lenovo machines and this is probably because that the Dell machine has larger wayness and size of LLC (see Table~\ref{tab:config}). 
For each machine on each setting, the number of eviction sets in each pool is almost the same as the LLC set-number, making the efficiency of selecting an eviction set comparable (within 290 milliseconds).

\subsubsection{Respective Minimal Eviction Set Selection}
TLB eviction set selection relies on a complete reverse-engineered mapping between virtual addresses and TLB sets~\cite{gras2018translation}, and thus it introduces no false positives, indicating that \upperpt can always select a matching eviction set for TLB).

However, selecting an LLC eviction set is based on profiling the access latency to a target address, described in Algorithm~\ref{alg:cache_flush}. As such, the profiled latency is not completely precise due to noise (e.g., interrupts) and may introduce false positives to the selection. To this end, we develop a kernel module to obtain the physical address of each L1PTE and thus verify whether the L1PTE is congruent with the eviction set selected by Algorithm 2. 
The experimental results show that the eviction set selection for LLC has no more than 6\% false positives in each system setting on each test machine. 

Note that selecting a TLB-based eviction set is within 1 microsecond while the LLC eviction set selection is within 290 milliseconds. Both of them are quite efficient, indicating that we can quickly start double-sided PThammering as mentioned below. 

\begin{figure}
\centering
\includegraphics[width=\columnwidth]{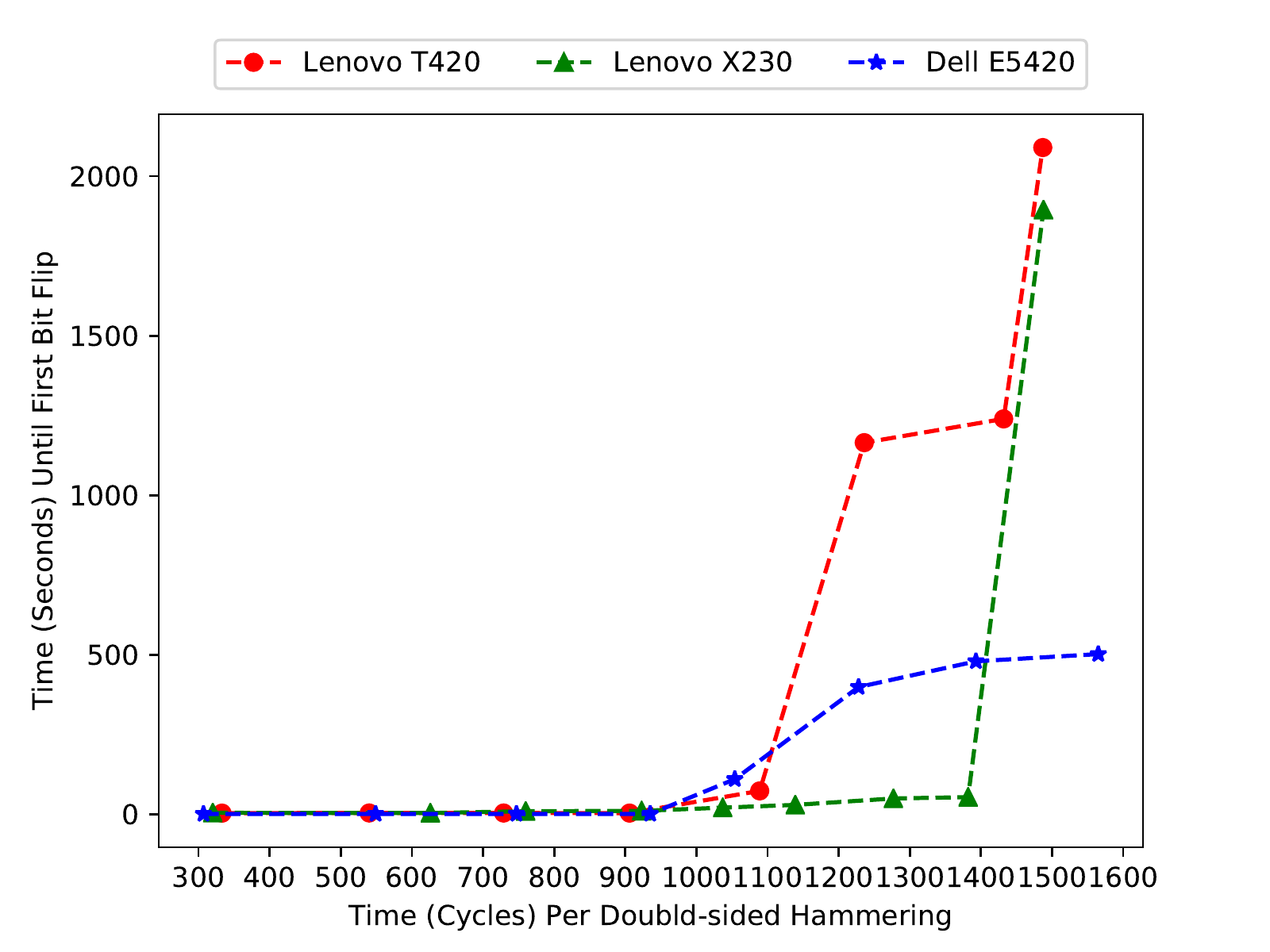}
\caption{As the time cost per double-sided hammering increases, the time to find the first bit flip also grows. When the time cost per hammering is greater than 1500 cycles in both Lenovo machines and 1600 cycles in the Dell machine, no bit flip is observed within 2 hours.} 
\label{fig:distribution}
\end{figure}

\subsubsection{Double-sided PThammering}\label{sec:double-sided pthammer}
To efficiently induce bit flips, we should hammer two L1PTEs that are one row apart within the same bank, similar to the way how double-sided hammering works. 
As such, we expect to select appropriate user virtual addresses such that their relevant L1PTEs meet the above requirement. 
However, the physical address of each L1PTE is required to know its location (i.e., DIMM, rank, bank, and row) in DRAM given that a physical address to a DRAM location has been reverse-engineering~\cite{xiao2016one,pessl2016drama}.
As we have no permission to access the kernel space, we cannot know the the physical address of a L1PTE, making it challenging to conduct double-sided PThammering.

To address this problem, we are inspired by previous works~\cite{cheng2018still,kwong2020rambleed,van2016drammer} and 
have two following steps. 
\emph{Firstly}, we select a pair of addresses that their respective L1PTEs are highly likely to be two rows apart. 
As the DRAM row size per row index (i.e., $RowsSize$) is publicly available (e.g., $RowsSize$ is 256KiB in each test machine), we can abuse the buddy allocator to force a large enough allocation of Level-1 page tables (e.g., 2GiB out of the total 8GiB DRAM for Level-1 page-tables, 8K times as large as $RowsSize$). By doing so, most page-table pages are in consecutive physical-page order and will exhaust consecutive rows with a high probability. 
On top of that, one Level-1 page-table page of 4KiB maps 512 physical pages of 4KiB. 
To this end, we choose such pair of addresses that their address difference is : $2 \cdot RowsSize \cdot 512$ (e.g., the address difference is 256MiB in our test machines).

\emph{Secondly}, loading two L1PTEs residing in the same bank of different rows triggers the row-buffer conflict~\cite{moscibroda2007memory} and causes clearly higher latency, compared to loading two L1PTEs that are within different banks. To this end, we perform TLB and LLC flushes for a selected address pair to load two L1PTEs from DRAM, and then profile the access latency to the address pair. 
If the latency is no less than a predetermined threshold, then the two L1PTEs are believed to be in the same bank with one row apart and the corresponding address pair will thus be sent for hammering. 
We identify the threshold at an offline stage by profiling access latency to 1000 address pairs of two kinds (i.e., one kind is that two L1PTEs for one address pair are in different rows of the same bank while the other one is that the two L1PTEs are in different banks). The experimental results show that each machine in both system settings has the same threshold. In each setting of each machine, no less than 95\% of address pairs are in the same bank when their access latency satisfies the threshold. Within the 95\% address pairs, no less than 90\% address pairs are in the same bank with one row apart. 

\subsubsection{Time Costs for \upperpt}\label{sec:timecosts}
As talked in Section~\ref{sec:pthammer}, the time cost per double-sided hammering must be no greater than the maximum latency allowed to induce bit flips. We firstly identify the maximum time cost that permits a bit flip on each machine through a published double-sided hammering tool~\footnote{
https://github.com/google/rowhammer-test}. 

The tool embeds two \texttt{clflush} instructions inside each round of double-sided hammering.
To increase the time cost for each round of hammering, we add a certain number of \texttt{NOP} instructions preceding the \texttt{clflush} instructions in each run of the tool. We incrementally add the \texttt{NOP} number so that the time cost per hammering will grow after each run. The time cost for the first bit flip to occur on each machine is shown in Figure~\ref{fig:distribution}. As shown in the Figure, 
the time cost until the first bit flip increases with an increasing cost per hammering. 
When the time cost per hammering is more than 1500 cycles on both Lenovo machines while 1600 on the Dell machine, not a single bit flip is observed within 2 hours. As such, 1500 and 1600 can be the maximum cost permitted to flip bits for the Lenovo and Dell machines, respectively. 

We then check whether the time taken by each round of double-sided PThammering is no greater than the permitted cost. For each double-sided PThammering, it requires accesses to two user virtual addresses as well as their respective TLB eviction set (i.e., 24 virtual addresses in total on each machine) and LLC eviction set (i.e., 26 virtual addresses on each Lenovo machine and 34 virtual addresses on the Dell machine). In both system settings, we conduct double-sided PThammering for 50 rounds and measure the time that each round takes. The experimental results (see Figure~\ref{fig:cyclerange} in Appendix~\ref{sec:appendix}) show that the time taken per double-sided PThammering is well below the aforementioned maximum cost on each machine, meaning that most address accesses within each PThammering are served by CPU caches rather than DRAM.  

As a result, we conduct \upperpt for 5 runs in both system settings and display the averaged time costs in Table~\ref{tab:timecosts}. The table clearly shows that we can successfully observe the first bit flip within 15 minutes of double-sided PThammering.


\subsection{Defeat the state-of-the-art software-only defenses}
To defend against rowhammer attacks, numerous software-only defenses have been proposed. Among the software-based defenses, CATT~\cite{brasser17can}, RIP-RH~\cite{bock2019rip} and CTA~\cite{wu2018CAT} are practical to mitigate existing rowhammer attacks in bare-metal systems. Note that RIP-RH~\cite{bock2019rip} enforces DRAM-based process isolation and thus prevents attackers from hammering target user processes. However, it does not protect the kernel and its page tables. Clearly, \upperpt can defeat it by inducing rowhammer bit flips in a L1 PTE and gain kernel privilege. 
In this section, we use Lenovo T420 to demonstrate proof-of-concept attacks against CATT~\cite{brasser17can} and CTA~\cite{wu2018CAT} respectively in the default system setting. 

\mypara{Compromise CATT~\cite{brasser17can}}
CATT~\cite{brasser17can} partitions each DRAM bank into a kernel part and a user part. These two parts are separated by at lease one unused row. When physical memory request is initiated, CATT allocates memory from either the kernel part or the user part according to the intended use of the memory. By doing so, CATT can confine bit-flips induced by the user domain to its own partition and thereby eliminate exploitable hammer rows that badly affect the kernel domain, the so-called physical kernel isolation.

However, we are still able to implicitly hammer kernel memory from the user domain by leveraging \upperpt. 
Specifically, our exploit has the following four phases:
\begin{enumerate}[label=\arabic*), noitemsep, topsep=2pt, partopsep=0pt,leftmargin=0.4cm]
\item Rely on the past work~\cite{cheng2018still} to allocate consecutive DRAM rows for Level-1 page-table pages (2GiB memory is allocated out of the total 8GiB DRAM);
\item Perform double-sided PThammering.
\item Verify whether ``exploitable'' bit flips have occurred by checking if a virtual address points to a page-aligned L1PTE. If not, go to step 2 and restart \upperpt (On average, 4 \upperpt-induced bit flips are needed to have one exploitable one);
\item If yes, we have gained the kernel privilege and we can gain the root privilege by changing \texttt{uid} of current process to \texttt{0}.
\end{enumerate}


\mypara{Compromise CTA~\cite{wu2018CAT}}
CTA (i.e., Cell-Type-Aware)~\cite{wu2018CAT} focuses on PTE-based privilege escalation rowhammer attacks. In such attacks, all the attackers induce bit-flips in L1PTEs such that the induced PTEs no longer point to the attackers' memory pages but instead point to other page-table pages of the same process, thereby gaining illegal access to the page tables. In order to destroy this core property, CTA  places Level-1 page tables in DRAM true-cells above a
``Low Water Mark" in the physical memory. If a PTE has a bit-flip in its physical frame number, it only points to a physical address lower than the ``Low Water Mark" rather than the page-table region. 

By leveraging \upperpt, we can break CTA and gain  root privilege. 
The key steps for the exploit are listed below:

\begin{enumerate}[label=\arabic*), noitemsep, topsep=2pt, partopsep=0pt,leftmargin=0.4cm]
\item We spray the physical memory under the ``Low Water Mark" with a large enough number of security critical structures, i.e., \texttt{cred} (note that \texttt{cred} stores the \texttt{uid} field.). Specifically, the attack process creates 32K child processes by invoking the {\tt fork} system call. For each child process creation, the kernel allocates a kernel stack and multiple  kernel structures including \texttt{cred}.  

\item Inside each child process, it firstly registers a signal and then goes to sleep. The registered signal will help the attack process wake up the child process when necessary. 

\item After completing the child-process creations, the attack process starts to occupy consecutive DRAM rows above the ``Low Water Mark" by forcing page-table page allocations (4GiB memory is used out of the total 8GiB DRAM). 

\item The attack process performs double-sided PThammering;
 
\item The attack process verifies whether ``exploitable'' bit flips have occurred by checking if a virtual address (VA) points to \texttt{cred} structure page. As the \texttt{cred} contains three user ids (e.g., \texttt{uid} and \texttt{suid}) and three group ids (e.g., \texttt{gid} and \texttt{sgid}) stored sequentially, the attack process can construct a unique string of the six ids and compare the string to the VA-pointing page. If the pointed page does not contain the string, then go to the step 4 to restart \upperpt (On average, 8 \upperpt-induced bit flips are needed to have one exploitable one);
 
\item If yes, the attack process has located a \texttt{cred} structure, changes \texttt{uid} to \texttt{0} and then wakes up every child process by delivering the registered signal. Inside the signal-catching function, each child process can check whether it has become a root process by invoking {\tt getuid}.
\end{enumerate}


%% file: discuss.tex
\section{Discussion}\label{sec:dis}

\mypara{Defeat ZebRAM~\cite{konoth2018zebram}}
ZebRAM is a rowhammer defense but only works for a virtualized system. We can extend \upperpt to defeat it in our future work.

Empirically, ZebRAM  observes that hammering a $row_i$ can only affect  adjacent $row_{i+1}$ and $row_{i-1}$. Based on this observation, ZebRAM leverages the hypervisor to split memory of a VM into safe and unsafe regions using even and odd rows in a zebra pattern. That is, all even rows of the VM are for the safe region that contains data, while all odd rows are for the unsafe region as swap space. As such, a rowhammer attack from the safe region can only incur useless bit flips in the unsafe region. For a rowhammer attack from the unsafe region, it is not possible since the unsafe region is inaccessible to an unprivileged attacker.   

However, our experimental results in Lenovo X230 show that $row_{i+2}$ and $row_{i-2}$ are able to induce bit flips in $row_{i}$, to be specific, 173 bit flips have occurred in $row_{i}$ within 16 hours. 
Clearly, ZebRAM's observation does not hold at least in our test machine and thus enabling both \upperperi and \uppertele to defeat ZebRAM in such a machine. 
Also, Kim et al.~\cite{kim2014flipping} report that ZebRAM's observation is not correct, i.e., hammering a row can affect three rows or more in a certain number of DRAM modules.

For those modules that support ZebRAM's empirical observation, an attacker can compromise ZebRAM as follows. ZebRAM does not protect the physical memory of the hypervisor and thus extended page tables (EPTs) residing in the hypervisor space are adjacent to each other. As such, an unprivileged attacker can initiate regular memory accesses to conduct \upperpt-like attacks, causing exploitable bit flips in EPT entries and escaping the VM. 


\mypara{Other Possible Instances of \uppertele}
Besides \upperpt, there might also exist other instances of \uppertele that leverage other built-in features of modern hardware/software. 
Particularly, features that focus more on functionality and performance may become potential candidates.  
For the hardware, we discuss about two famous CPU features. Specifically, out-of-order and speculative execution are two optimization features that allow a parallel execution of multiple instructions to make use of instruction cycles efficient. 
As such, an unprivileged attacker can leverage such hardware features to bypass user-kernel privilege boundary and access kernel memory ~\cite{kocher2018spectre,lipp2018meltdown}.

For the software, we talk about OS kernel features that handle local and network requests. 
A system call is a programmatic feature in which a user application requests a service from the kernel. By invoking a system call handler, a user indirectly accesses the kernel memory.
A network I/O mechanism is also a programmatic feature that allows the OS to serve requests from the network. Particularly, the network interface card (NIC) will throw out a hardware exception to notify the kernel of each network packet NIC receives. Within the exception handler, the kernel will access kernel memory. Thus, a remote user invokes this feature to access kernel memory. 

As a result, an attacker might build up an exploitable communication path to a target kernel address by abusing the above features. 

\mypara{Mitigation}
we might detect both \uppertele and \upperperi by performance counters~\cite{aweke2016anvil}. However, such anomaly-based detection is prone to false positives and/or false negatives by nature~\cite{brasser17can}.

Alternatively, we might take hardware defenses such as PARA~\cite{kim2014flipping}, TRR~\cite{DDR4, LPddr4} and TWiCe~\cite{lee2019twice} to increase DRAM refresh rate for specified rows, which would reduce ${\mathtt T_{max}}$ in Definition~\ref{def:telehammer} (see section~\ref{sec:telehammer}) as much as possible so as to break the last time condition in the definition. Unfortunately, they require new hardware designs and thus cannot be used to protect legacy systems.   

For \upperpt, we might cache PTEs in an isolated cache to eliminate a communication path identified by \upperpt. Since PTEs are placed in a separated cache, then \upperpt cannot use the cache-eviction approach to evict PTEs. However, reserving an isolated cache only for page-table pages is expensive in hardware and requires re-designing hardware.
Even if such an isolated cache for PTEs would be released by CPU manufacturers, there might exist other communication paths for \upperpt to hammer PTEs, or other instances of \uppertele that hammers other critical structures in the kernel space. 
Summarizing, we believe that \uppertele-based rowhammer attacks are hard to be mitigated. 

%% file: conclusion.tex
\section{Conclusion}\label{sec:conclusion}
In this paper, we first observed a critical condition required by existing rowhammer exploits to gain
the privilege escalation or steal the private data. We then proposed a new class of rowhammer attacks, called \uppertele, that crosses privilege boundary and thus eschews the condition. Besides, we presented a formal model to define key conditions to set up \uppertele and \upperperi and summarized three advantages of TeleHammer over \upperperi. 

On top of that, we created an instance of \uppertele, called \upperpt that can cross the user-kernel boundary and induce bit flips in Level-1 page table entries. Our experimental results on three test machines showed that the first cross-boundary bit flip occurred within 15 minutes of double-sided PThammering. Furthermore, we developed \upperpt-based attacks that allow an unprivileged attacker to compromise the state-of-the-art software-only defenses in default system setting. 

%% file: appendix.tex
\section{Time cost per double-sided PThammering}\label{sec:appendix}

\begin{figure*}
	\begin{subfigure}{\columnwidth}
		\includegraphics[scale=0.5]{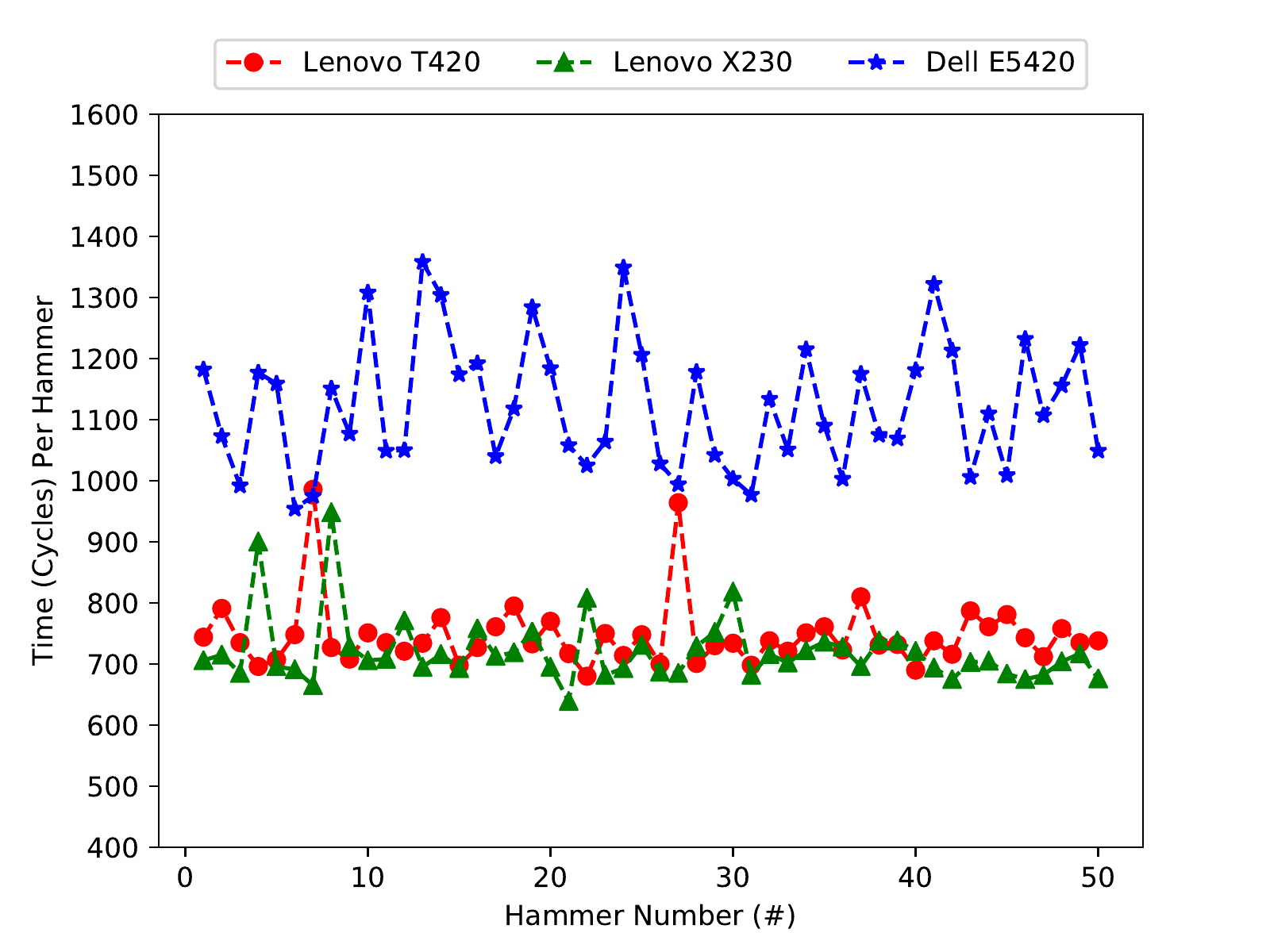}
		\caption{Double-sided PThammering in the default \emph{regularpage} setting.}
		\label{fig:4kbcyclerange}
	\end{subfigure}
	\hfill
	\begin{subfigure}{\columnwidth}
		\includegraphics[scale=0.5]{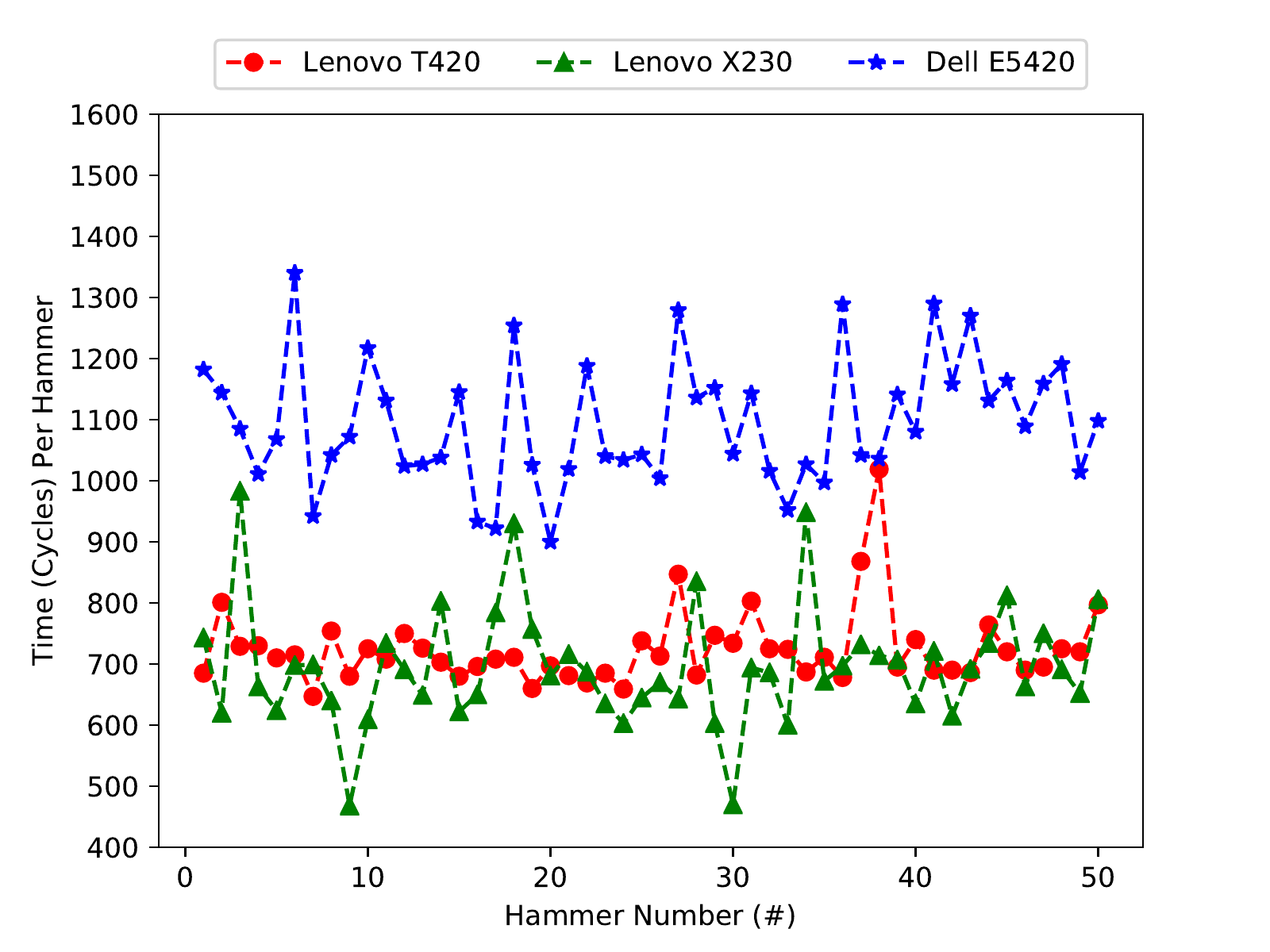}
		\caption{Double-sided PThammering in the  \emph{superpage} setting.}
		\label{fig:2MBcyclerange}
	\end{subfigure}%
	\caption{In both system settings, the time-cost range on each machine is well below the maximum time cost (see Figure~\ref{fig:distribution}) that allows bit flips.}
	\label{fig:cyclerange}
\end{figure*}

In either system setting, we conduct double-sided PThammering for 50 rounds on each machine and measure the time that each PThammering takes, shown in Figure~\ref{fig:cyclerange}.

As Figure~\ref{fig:4kbcyclerange} of the \emph{regularpage} setting shows, most time costs per double-sided PThammering (no less than 96\%) in both Lenovo machines are in the range of \{600, 900\} (100\% time costs are below 1000 cycles) while the Dell machine has a range of \{900, 1400\}.

In Figure~\ref{fig:2MBcyclerange} of the \emph{superpage} setting, most time costs per double-sided PThammering (no less than 94\%) in both Lenovo machines are in the range of \{400, 900\} (100\% time costs are below 1100 cycles) while the Dell machine has the same range of \{900, 1400\}.

Given the maximum time cost that allows bit flips in Figure~\ref{fig:distribution} of Section~\ref{sec:timecosts}, we conclude that \upperpt is efficient enough to induce bit flips.

